\documentclass[showpacs,preprintnumbers,pre,floatfix]{revtex4}
\usepackage{epsf}
\usepackage{graphicx}
\usepackage{dcolumn}% Align table columns on decimal point
\usepackage{bm}
\begin{document}
%\draft
\title{Helical magnetorotational instability in 
a Taylor-Couette flow with strongly reduced Ekman pumping}

\author{Frank Stefani$^1$, Gunter Gerbeth$^1$, Thomas Gundrum$^1$, Rainer Hollerbach$^2$, J\={a}nis Priede$^3$, G\"unther R\"udiger$^4$, and Jacek Szklarski$^{1,5}$}
\affiliation{$^1$Forschungszentrum Rossendorf, 
P.O. Box 510119, D-01314 Dresden, Germany\\
$^2$Department of Applied Mathematics, University
of Leeds, Leeds, LS2 9JT, United Kingdom\\
$^3$Coventry University, Priory Street, Coventry CV1 5FB, 
United Kingdom\\
$^4$Astrophysikalisches Institut Potsdam, 
An der Sternwarte 16, D-14482 Potsdam, Germany\\
$^5$IPPT, ul. \'{S}wi\c{e}tokrzyska 21,
00-049 Warsaw, Poland}

%\date{Submitted to Phys. Rev. Lett.}

\begin{abstract}
The magnetorotational instability (MRI) is thought to play a key role
in the formation of stars and black holes by sustaining the turbulence
in hydrodynamically stable Keplerian accretion discs. In previous
experiments the MRI was observed in a liquid metal Taylor-Couette flow
at moderate Reynolds numbers by applying a helical magnetic field.
The observation of this helical MRI (HMRI) was interfered with a 
significant
Ekman pumping driven by solid end-caps that confined the instability
only to a part of the Taylor-Couette cell. This paper describes the
observation of the HMRI in an improved Taylor-Couette setup with
the Ekman pumping significantly reduced by using split end-caps. The
HMRI, which now spreads over the whole height of the cell, appears
much sharper and in better agreement with numerical predictions. By
analyzing various parameter dependencies we conclude that the observed
HMRI represents a self-sustained global instability rather than a
noise-sustained convective one. 
\end{abstract}


\maketitle
%\narrowtext

\section{Introduction}

Cosmic magnetic fields are produced, on a wide range of
spatial scales, by the homogeneous 
dynamo effect in electrically conducting fluids 
\cite{RUHOBUCH}. The 
Earth's  magnetic field is generated in its outer
core, most likely by spiral flow 
structures which result from 
the combined action of Coriolis forces and 
thermal and/or compositional buoyancy \cite{EARTH}.
The magnetic field of the Sun is generated 
in the convection zone and in the tachocline, very likely 
under the influence of differential rotation and 
helical turbulence \cite{SUN}.
A similar mechanism is probably at the root of 
large scale galactic magnetic fields \cite{GAL}, 
while inter-galactic magnetic fields
are thought to be a product of the so-called fluctuation 
dynamo \cite{FLUCT}.

Magnetic fields play an active role 
in cosmic structure formation.
It was in 1991 that Balbus and Hawley \cite{BAHA} highlighted the 
importance of the magnetorotational instability (MRI,               %!!!
or Velikhov-Chandrasekhar instability \cite{VELIKHOV,CHANDRA1})
for the angular momentum 
transport in accretion disks around stars and black holes.
In MRI, the externally applied magnetic field serves 
only as a trigger 
for the instability that actually taps into the rotational energy 
of the flow. This is quite in contrast to another magnetic 
instability    in which prevailing currents 
in the fluid can become unstable by 
themselves.
This latter, so-called Tayler instability \cite{TAYLER} 
is held responsible for parts of the dynamo mechanism in 
stars \cite{SPRUIT1}, 
and for some observed helical structures in jets 
and outflows \cite{SPRUIT2}.

Significant theoretical and computational progress has been  %!!!
made in the understanding of dynamo action and 
magnetic instabilities.
Besides this, the last decade has also seen considerable  
experimental activities
to investigate those effects in the liquid metal laboratory 
\cite{RMP,ZAMM}.
At the end of 1999, the hydromagnetic dynamo effect was 
observed nearly simultaneously
at the large-scale liquid sodium facilities in 
Riga \cite{PRLRIGA} and Karslruhe \cite{KARLSRUHE}. 
In 2006, the VKS-experiment in Cadarache has shown 
not only self-excitation \cite{MONCHAUX} but
also impressive magnetic field reversals \cite{BERHANU}, 
although the use
of soft-iron impellers with high relative permeability 
makes the interpretation of these results cumbersome. 

As for the so-called ``standard MRI'' (SMRI),  
with only an axial magnetic field applied, 
the authors of \cite{LATHROP} claim to 
have observed it in a spherical Couette flow, but their
background state was already fully turbulent, thereby 
defeating the
original idea that the MRI would destabilize an 
otherwise stable flow.

There is a simple reason for the flow 
being turbulent before the SMRI sets in. The point          %!!!  %!!! Thomas
is that the {azimuthal magnetic field component             %!!!
of the MRI mode can only  be produced from the axial field  %!!!
by induction effects, which are proportional to the magnetic 
Reynolds number ($Rm$) of the flow. $Rm$, in turn, is               %%%%%
proportional to the hydrodynamic Reynolds number according to 
$Rm=Pm Re$, where the magnetic Prandtl number $Pm=\nu/\lambda$
is the ratio of viscosity $\nu$ to magnetic 
diffusivity $\lambda=1/\mu_o \sigma$. 
For liquid metals $Pm$ is typically in the range $10^{-6}...10^{-5}$.
Therefore, in order to achieve $Rm \sim 1$, we need 
$Re\sim 10^5...10^6$, and wall-constrained flows 
with such high $Re$ are known to be turbulent, whatever
the linear stability analysis might tell. Apparently 
in contrast with this
common wisdom, a 
group in Princeton has recently shown \cite{BURIN} that
a flow profile in a short Taylor-Couette cell 
can be kept stable until $Re \sim 10^6$ if
an appropriate configuration of split end-caps is used.
However,  the mechanisms that ultimately leads 
to flow stabilization
in this configuration are not well understood. 
At least, the optimum rotation ratios of the 
split rings do not agree with numerical predictions, a
discrepancy that might be explained by some wobbling of the inner 
cylinder  \cite{ROACH}.

Having learned that 
$Rm \sim 1$ (and hence $Re \sim 10^6$) 
is needed just to
produce the azimuthal magnetic field component of the MRI mode
from the applied 
axial field, one might ask why not  
substitute this induction process by simply 
{\it externally applying an azimuthal
magnetic field as well\/}? This idea was pursued in 
\cite{HORU,HORUAN}, where it was shown that this ``helical MRI'' (HMRI), 
as we now call it, 
is indeed possible at
far smaller Reynolds numbers and 
magnetic field amplitudes 
than SMRI.  

In \cite{HORU} it was also shown 
that  HMRI and SMRI are continuously connected. %!!!
In Fig.\ 1     of \cite{HORU} the critical
$Rm$ was plotted against the rotation 
ratio $\mu:=f_{\rm out}/f_{\rm in}$
of outer to inner cylinder of a Taylor-Couette set-up.
The extremely steep increase of this curve
at the Rayleigh line, which occurs for a purely axial 
magnetic field, is just smeared out when an azimuthal 
field is added. Although HMRI, in the limit of 
small $Re$,  
is a weakly destabilized inertial oscillation 
\cite{LIU1}, there is a 
continuous and monotonic transition to SMRI 
when $Re$ and the magnetic field 
strength are 
increased simultaneously.

The relation of HMRI, which appears as a 
travelling wave, and SMRI 
is currently the subject 
of intense discussions in the 
literature 
\cite{PRIEDE1,JACEK1,LIU2,JACEK2,JACEK3,LAKHIN,JACEK4,LIU3,PRIEDE2},
the roots of which trace back to an 
early dispute between Knobloch
\cite{KNOB1,KNOB2} and Hawley and Balbus \cite{HAWLEYBALBUS}.

A remarkable property of HMRI for small $Pm$ (which has been
coined ``inductionless MRI''), was
clearly worked out in
\cite{PRIEDE1}. It is the apparent paradox that a
magnetic field is able to trigger an instability 
although the total energy dissipation of the system is 
larger than without this field. 
This is not so surprising though
when seen in the context of 
other {\it dissipation induced instabilities} which 
are quite common in many areas of physics \cite{RMPDISSIPATION}.

Another, and not completely resolved issue concerns the relevance
of HMRI for astrophysical flows. On first glance, HMRI seems very 
attractive as it could extend the range of applicability of MRI 
into those regions of accretion disks 
which are characterized by a small $Pm$.
This might apply  to the ``dead zones'' 
of protoplanetary disks \cite{DEADZONES}
as well as to the outer parts of accretion 
disks around black holes \cite{BALBUSPM}. 

However, before discussing this point in more detail, 
it has to be checked
whether HMRI works at all for Keplerian rotation profiles 
$\Omega(r)\sim r^{-3/2}$  or not. 
In the inductionless limit, i.e. for $Pm=0$, 
and using a local WKB analysis in the small-gap approximation, 
the answer to 
this question is ``no''  \cite{LIU1}. In contrast to this
the solution of the 
corresponding global eigenvalue equation gives an
affirmative answer,
as long as at least the outer or the 
inner radial boundary is 
electrically conducting
\cite{HOLLERBACHRUEDUGERPRE}. This is certainly 
good news for the working of HMRI in the outer, 
cold  parts of accretion 
disks since
their inner parts  have indeed a higher 
conductivity.

Unfortunately, even this is not the end of the story. 
Both methods, the WKB method  and the global
eigenvalue equation, deal with waves with a single
wavenumber and frequency. However, since HMRI appears 
in the form of a travelling wave, one has to be careful with 
the interpretation of the positive growth rate of a single 
monochromatic wave. The crucial point here is that
those monochromatic waves are typically not able
to fulfill the axial boundary conditions at the
ends of the considered region.
To fulfill them,  one has to consider wave packets. 
Only wave packets with vanishing group velocity 
will remain in the finite length system. 
Typically, the onset of this {\it absolute instability}, 
characterized by
a zero growth rate {\it and} a zero group velocity, is harder 
to achieve  than the 
{\it convective instability} of a monochromatic wave 
with zero growth rate. 
A comprehensive analysis of the relation of convective, absolute,
and global instability can be found in the survey papers
\cite{HUERRE,CHOMAZ}. In the context of MHD, 
this distinction played an important role, e.g.,
in the Riga dynamo experiment \cite{GAILITISFREIBERGS}.
A detailed 
analysis of the relation of convective and 
absolute instability for HMRI
can be found in \cite{PRIEDE2}. 
From the extrapolation of the results
of this paper it seems that 
Keplerian rotation profiles are indeed 
absolutely HMRI-unstable, but a final solution to
this puzzle is still elusive.

Notwithstanding these ongoing debates, 
the dramatic decrease of the
critical rotation rates and magnetic field intensities 
for HMRI, as compared with SMRI, 
made this new type of MRI 
particularly attractive for experimental studies. 

First experimental evidence for HMRI 
was obtained in 2006 at the liquid metal facility 
PROMISE ({\it P\/}otsdam {\it RO\/}ssendorf
{\it M\/}agnetic {\it I\/}n{\it S\/}tability {\it E\/}xperiment)
which is basically  a Taylor-Couette (TC) cell made of 
concentric rotating 
copper walls, filled with  GaInSn (a eutectic
which is liquid at room temperatures).
In \cite{PRL} it was shown that the
MRI travelling wave appears only in 
the predicted finite window of the magnetic field intensity, 
with a 
frequency of the
travelling wave that was also in good accordance with 
numerical simulations. More detailed results were 
published in \cite{APJL,NJP}. 

However, a slight difficulty with these early experiments was the fact
that the MRI wave did not travel along the entire height of the
Taylor-Couette cell but ceased to exist at some position typically 
close  to mid-height. By modifying slightly the electric boundary 
conditions in radial direction, and analyzing the resulting 
change of the   
axial velocity \cite{AN}, it was possible
to identify this sink of the 
travelling wave with 
the position of the radial jet 
that originates from the Ekman pumping at 
the upper and lower lids of the TC cell \cite{KAGEYAMA}.
The influence of this Ekman pumping on the flow structure 
was extensively 
discussed in a number of papers \cite{JACEK1,JACEK2,JACEK3}.

A related critical point is connected with electric currents
which are induced in the lower end-cap that was made of copper 
in the first PROMISE experiments (now called PROMISE 1). 
In the worst case these
currents 
are able to drive a significant Dean flow which possibly 
could modify the original Taylor-Couette flow profile 
\cite{JACEK2,PRIEDE3}.

In order to overcome both problems we have changed 
the axial boundary conditions, first by
replacing the copper of the lower lid by insulating material, 
and second 
by splitting  the (upper and  lower) lids into
two rings
(rotating with the inner and outer cylinder, respectively)
in order to minimize the 
Ekman pumping. 
Henceforth, this new version of the experiment
will be  called 
PROMISE 2 in order to discriminate it from the 
previous PROMISE 1 experiment. 

In this paper we will show that the minimization of 
the Ekman pumping makes the 
radial jet disappear almost completely. 
Consequently, the MRI wave can travel throughout  
the entire TC cell. We will investigate in detail four
parameter dependencies of the MRI wave 
showing a much improved agreement
with the 2D numerical predictions. Further comparisons 
with  the results of a  1D eigenvalue code for
the convective and absolute instability 
will provide strong evidence that the observed wave represents a
global instability and not only a noise-triggered convective 
instability as argued in \cite{LIU2,LIU3}.

\section{From PROMISE 1 to PROMISE 2}

Apart from the modified end-caps,
the PROMISE 2 set-up is  more or less identical to 
that of PROMISE 1 which was described in detail in 
\cite{PRL,NJP}.
Its main part is a cylindrical vessel 
(cp. Fig.\ \ref{set-up1}) made of 
two concentric copper cylinders. 
The inner one is 10 mm thick
extending in radius from 22 to 32 mm; the outer wall is 15 mm thick,
extending from 80 to 95 mm in $r$.  This vessel is 
filled with the eutectic 
alloy Ga$^{67}$In$^{20.5}$Sn$^{12.5}$ whose physical properties
at 25$^{\circ}$C are as follows: density
$\rho=6.36 \times 10^3$ kg/m$^3$, kinematic viscosity $\nu=3.40\times
10^{-7}$ m$^2$/s, electrical conductivity $\sigma=3.27\times 10^6$
($\Omega$ m)$^{-1}$. The magnetic Prandtl number is then $Pm=\mu_0
\sigma \nu=1.40 \times 10^{-6}$.

The copper vessel is closed by a plastic bottom and is
fixed, via an aluminum spacer, on a precision 
turntable; the outer wall of the vessel thus serves as the outer
cylinder of the TC cell. The inner cylinder of the TC flow is 
fixed to an upper turntable, and is immersed into the 
liquid metal from above.  It has
a thickness of 4 mm, extending in 
radius from 36 to 40 mm.  Between this immersed cylinder and
the inner wall of the containment vessel there is thus
a gap of 4 mm, also filled with GaInSn.
The actual TC cell then extends in radial direction 
over a cylindrical
gap of width $d=r_{\rm out} - r_{\rm in}=$ 40 mm from the outer 
surface of the inner
cylinder at $r_{\rm in}=40$ mm to the inner surface of the outer 
wall of the
containment vessel at $r_{\rm out}=80$ mm, and in axial direction 
over the
liquid metal height of 0 mm $\le z \le 400$ mm.

While in PROMISE 1 the upper end-cap was a
plastic lid fixed to the outer frame this 
end-cap is now split 
into two concentric plastic rings (see Fig.\ \ref{set-up2}).
The inner ring rotates with the inner
copper cylinder, the outer one rotates with the outer
cylinder. By this means, the opposing Ekman pumping 
effects at the rings
compensate each other to a large extent.
The choice of the splitting position at $r=56$ mm 
was motivated by the simulation in
\cite{JACEK2} showing that a splitting at $0.4 \; d$ 
would minimize the
global Ekman pumping for the case with the                    %!!! Jacek
two rings attached to the cylinders (using only two
rings significantly simplifies the construction  when 
compared to a many-rings case).
The lower end-cap is split in the same manner, 
so that we have a
symmetric configuration, with respect to both
the rotation rate and to the electrical conductivities. 
This is a significant improvement compared with 
PROMISE 1, in which the copper bottom was simply 
part of the vessel which rotated with the outer cylinder.

The magnetic field configuration is identical to 
that of PROMISE 1, with axial magnetic fields of 
order 10 mT being produced by the current $I_{\rm coil}$ 
in a double-layer 
coil with 78 windings, and
an azimuthal field of the same order being generated 
by a current $I_{\rm rod}$ through a central water-cooled 
copper rod of radius 15 mm.

As in PROMISE 1, the measuring instrumentation 
consists exclusively of 
two high-focus ultrasonic transducers 
(from Signal Processing SA)
with a working frequency of 4 MHz
which are fixed into the outer plastic ring, 
12 mm away from the outer copper wall, flush
mounted at the interface to the GaInSn. 
Since this outer ring is rotating, it is necessary
to transfer the signal into the laboratory frame
(in contrast to PROMISE 1 in which the upper lid
was fixed to the frame).
This is accomplished by the use of a slip ring    %!!!
contact which is situated below the TC vessel
(not shown in Fig.\ \ref{set-up1}).

The advantage of the  Ultrasound Doppler
Velocimetry (UDV) is that it provides full 
profiles of the  axial velocity $v_z$ along the beam-lines 
parallel to the axis of rotation. The spatial resolution 
in axial direction is chosen 0.685 mm, the 
time resolution is 
1.84 sec. The width of the beam (over which $v_z$ 
is averaged) is approximately 8 mm,  according 
to the diameter of the ultrasonic transducers. 

To get reliable signals from along the 400 mm height 
of the TC cell one has to ensure a good 
acoustic contact between the transducer and the GaInSn
which is often compromised by non-perfect wetting, sticking 
oxides or bubbles. A good signal is all the more 
important since the 
velocity perturbations to  be measured are typically 
in the order of 0.1 mm/s, which 
is at the lower edge of the applicability
of the UDV technique.

This fact has to be kept in mind when comparing, i.e.,
rms values of the velocity perturbation
from different runs.
For example, the identification of the transition between
stable and unstable regimes rely strongly on
the measured velocity distribution. 
However, this velocity signal is typically affected by noise 
which, in turn, depends on the acoustic contact.

Typically, the duration of the experimental runs 
was 1900 sec, after a
waiting time of one hour or even longer in cases when 
long transients were to be expected.

\section{Results}

In this section, we will present the main experimental
results and compare them with numerical results 
based on a 1D eigenvalue solver (for details consult 
\cite{PRIEDE1,PRIEDE2}) 
and on a time-stepping 2D solver                       %!!! Jacek Fournier raus
that was already used in \cite{JACEK2}.
Usually, for both solvers perfectly conducting     
boundary conditions were applied at the inner and outer
radius, which is certainly not completely correct since
the conductivity of copper is approximately 
13 times higher than that of GaInSn, but not infinite.
For some parameters, at least, it has been checked that 
the inclusion of the finite-thickness copper walls 
in the 1D eigenvalue code 
does not change significantly  the results
based on perfectly conducting walls.

In particular we will study the dependence of the 
HMRI  on the four parameters that govern the problem.
Given the inner and outer radii of the TC cell,
$r_{\rm in}$ and $r_{\rm out}$ (with a gap width 
$d=r_{\rm out}-r_{\rm in}$), the rotation 
rate of inner and 
outer cylinder, $f_{\rm in}$ and $f_{\rm out}$, 
the axial magnetic 
field $B_z$ and and the azimuthal magnetic 
field at the
inner cylinder,  $B_{\varphi}(r=r_{\rm in})$,
these governing parameters are the Reynolds number
\begin{eqnarray}
Re:=2 \pi f_{\rm in} d/\nu \; ,
\end{eqnarray}
the ratio of rotation rates
\begin{eqnarray}
\mu:=f_{\rm out}/f_{\rm in} \; ,
\end{eqnarray}
the Hartmann number
\begin{eqnarray}
Ha:=B_{z} (r_{\rm in}
d \sigma/\rho \nu)^{1/2}  \; ,
\end{eqnarray}
and the ratio of azimuthal to axial magnetic field
\begin{eqnarray}
\beta:=B_{\varphi}(r=r_{\rm in})/B_{z} \; .
\end{eqnarray}

With the actual geometry and the material properties 
of 
GaInSn given, we can derive the following useful
relations between
the experimental parameters and the dimensionless 
parameters:
\begin{eqnarray}
Re=29586 \cdot f_{\rm in}/{\rm Hz} ,
\end{eqnarray}
\begin{eqnarray}
Ha=0.158 \cdot I_{\rm coil}/{\rm A},
\end{eqnarray}
\begin{eqnarray}
\beta=0.0491 \cdot I_{\rm rod}/I_{\rm coil}.
\end{eqnarray}

Before entering into the details of the 
parameter dependencies of the MRI, we present the
results in the magnetic field free case.

\subsection{Results without magnetic fields}

The flow structure in the non-magnetic case
is an important reference point for the later 
analysis of the MRI. Starting with $\mu=0$, we 
expect a Taylor vortex flow (TVF) whose amplitude 
decreases for increasing $\mu$. Beyond the Rayleigh point 
$\mu_{\rm Rayl}:=r^2_{\rm in}/r^2_{\rm out}=0.25$, the TVF is 
supposed to
disappear completely in accordance with Rayleigh's criterion.
However, in any real TC-cell with finite height
this is far from trivial due to the existence of a
 meridional flow driven by the Ekman pumping at the upper and 
lower lids. Approximately at mid-height of the TC cell the
two opposed vertical flows meet each other and  
produce a radial jet \cite{KAGEYAMA}.

The difference of the flows in the 
PROMISE 1 and PROMISE 2 configuration, respectively, 
is documented in Fig.\ \ref{withoutfield}. What is actually
shown is the velocity component 
measured along the ultrasonic beam-lines. 
This velocity component 
is supposed to represent 
the  total axial velocity $V_z(z,t)$, although some slight 
influence of 
the  strongly dominant $V_{\varphi}$ component cannot be 
ruled out in case
the direction of the UDV transducers is not
perfectly vertical. On the r.h.s. of 
each panel the time-averaged value $va_z(z)$ is
given in dependence on the axial position $z$.

Figure \ref{withoutfield} 
shows, at $f_{\rm in}=0.1$ Hz,  the results for PROMISE 2, 
(a), (b), and (c) corresponding to $\mu=0$, $\mu=0.1$, and $\mu=0.27$, 
respectively. As expected, we observe a TVF which is getting weaker
for increasing $\mu$ and which disappears completely at $\mu=0.27$.
For comparison, Fig.\ \ref{withoutfield}d shows one result
for PROMISE 1, at $f_{\rm in}=0.06$ Hz and $\mu=0.27$.
Here we observe a very steep sign change of 
$V_z(z)$ approximately at mid-height which necessarily must 
be connected with a radial jet at this point, a feature that
is nearly absent in PROMISE 2 (Fig.\ \ref{withoutfield}c). 
Evidently the global effect of the 
Ekman pumping is
drastically reduced by the splitting of the end-caps.      %!!!
In the following we will see that this absence of the
radial jet has important consequences for the 
propagation of the MRI wave.

\subsection{Variation of $Re$}
As with many other instabilities in hydrodynamics, MRI 
is expected to set in at a certain critical Reynolds 
number $Re_{\rm crit}$.
In order to study this transition in detail, 
we have fixed the rotation ratio to $\mu=0.27$ 
and the current through the
coil to 
$I_{\rm coil}=76$ A (i.e. $Ha=12$). 
The current through the central rod has been set to
either $I_{\rm rod}=4000$ A (i.e. $\beta=2.59$) or
$I_{\rm rod}=7000$ A (i.e. $\beta=4.53$).
Throughout the paper (if not otherwise stated)          %!!!
we have chosen the rotation to be clockwise,  
when viewed from above, and       %!!!
$B_z$ and $I_{{\rm rod}}$ directed downward, 
where the latter induces $B_{\varphi}$ in the clockwise
direction.                          %!!!

To start with, we show in Fig.\ \ref{individualsensors} the 
axial velocity signals, again  the time averaged 
$va_z(z)$ on the r.h.s, 
but here and henceforth the deviation 
$v_z(z,t):=V_z(z,t)-va_z(z)$ of the 
total velocity from the time averaged value
on the l.h.s.
The panels (a) and (b) exhibit the individual signals from the
UDV sensors 1 and 2, respectively, (c) shows 
their average, and (d) shows
their difference.
Actually, in this difference signal one observes a 
slight $m=1$ mode 
with a dominant frequency equal to the
rotation rate of the outer cylinder,
which
results just from some slight asymmetries of the
facility (e.g. the current leads to the central rods).
This is in 
contrast to PROMISE 1
where we have observed, in some parameter 
regions \cite{PRL,NJP,AN}, an
$m=1$ mode with a frequency between the rotation rate of inner 
and outer cylinder whose origin is not yet clarified.

For $I_{\rm rod}=7000$ A and $I_{\rm coil}=76$ A,
the 12 panels of 
Fig.\ \ref{Relinie} 
show $V_z(z,t)$ for increasing $f_{\rm in}$ (i.e. $Re$). 
From here on, the l.h.s. of each panel exhibits 
the average over the two sensors (corresponding  to panel 
(c) in Fig.\ \ref{individualsensors}).
Evidently, the travelling wave instability 
sets in approximately between 
$f_{\rm in}=0.04...0.05$ Hz (i.e. $Re=1183...1479$) and 
remains until the highest value $f_{\rm in}=0.24$ Hz 
(i.e. $Re=6830$) that we have studied in the experiment. 

It is also instructive to see what happens when we 
change the direction of $B_z$ by changing the direction 
of the current in the coil. It is well-known
that the direction of
propagation, i.e. whether the wave pattern drifts in 
the +$z$ or $-z$ direction,
depends on whether the product $\Omega B_z B_{\varphi}$ 
is negative or positive, respectively.
Figure \ref{direction} documents this
effect for the parameter choice $f_{\rm in}=0.1$ Hz,
$\mu=0.27$, $I_{\rm rod}=7000$ A, and
$I_{\rm coil}=\pm 76$ A. Note that the 
shape of the (a) upward and (b) downward travelling wave is now quite 
symmetric, 
in contrast
to PROMISE 1 in which the 
position of the radial jet made
the pattern more asymmetric (see, e.g, Fig.\ 4 in \cite{AN}).

When speaking about upward and downward travelling wave 
it is also necessary to say that these two 
types do not appear exclusively. Actually, the 
upward travelling wave, when reflected at the upper boundary,
produces a slight downward travelling wave, and vice versa.
To analyze this point we show 
in Fig.\ \ref{ReFFT} the 
2D Fourier transform from the $z,t$-space to the $k,f$-space of      %!!!
the signal $v_z(z,t)$ for $f_{\rm in}=0.05$ Hz. 
Not surprisingly, the
experimental data are more smeared out than 
the numerical predictions. Nevertheless, 
the position of the maximum in
the $k,f$  space is approximately the same in both cases.
The maximum of the 2D FFT is mainly localized in the second and 
fourth quadrant 
of the $k,f$-space which indicates  an 
upward travelling wave. However, the slight signal in the
first  and third quadrant indicates also the existence 
of a downward travelling wave.

By applying a Hilbert transform to the original data, 
it is possible 
to disentangle the dominant upward travelling wave and the
minor downward travelling wave which results 
from a reflection of the upward travelling wave 
at the upper end-cap.
This is done, now again for $f_{\rm in}=0.1$ Hz, 
$\mu=0.27$, in Fig.\ \ref{updown}.
Inspired by \cite{CHIFFAUDEL}, we will
later use the ratio of the maxima of the downward to the upward
travelling wave as an additional indicator for the onset of the
absolute instability.

In Fig.\ \ref{Revergleich} we restrict ourselves to the 
downward travelling wave and compare its shape
 with  simulated results for the particular cases 
$f_{\rm in}=0.04$ Hz and $f_{\rm in}=0.05$ Hz.
We observe a quite convincing agreement of 
experimental data and numerical simulation, even the 
amazing slow decay of a  transient wave at  
$f_{\rm in}=0.04$ Hz is recovered by the numerics.

A more quantitative comparison  of various experimentally 
and numerically obtained features of the MRI wave, 
in dependence on $f_{\rm in}$,  
is made  in Fig.\ \ref{Reanalyse}.
This figure contains four parts: Fig.\ \ref{Reanalyse}a shows 
the squared rms of $v_z(z,t)$, calculated between 250 mm$<z<$350 mm 
(this particular range has been chosen since the velocity in this 
interval is 
comparably high and the UDV signal is usually not 
much affected by noise as it is
for smaller values of $z$).
Figure \ref{Reanalyse}b
shows the squared ratio of the maximum amplitude 
(taken over all $z$) of the
downward to that of  the upward travelling wave. This quantity had 
been
used successfully in other problems 
to distinguish between absolute and
convective instability \cite{CHIFFAUDEL}. 
Figures \ref{Reanalyse}c and 
\ref{Reanalyse}d
show the normalized wave frequency $f_{\rm wave}/f_{\rm in}$ 
and the normalized phase velocity of the wave 
$c_{\rm wave}/(f_{\rm in} r_{\rm in})$, 
respectively.
The same dependencies, but 
now in more detail for 0.02 Hz $<f_{\rm in}<$ 0.1 Hz are
shown in Fig.\ \ref{Reanalyse_detail}.

Basically, the agreement between experiment and numerical results
from the 2D code 
is quite convincing,  
in particular the onset of the instability approximately 
at $f_{\rm in}=0.4$ Hz for $I_{\rm rod}=7000$ A.
However, there is one significant deviation: 
while the numerics 
indicate a disappearance of the instability approximately 
at $f_{\rm in}=0.11$ Hz,
the experiment shows a continuation of the instability far
beyond this point. The reason of this difference is not
yet clear. What can be said, at least, is that the chosen
parameter $\mu=0.27$ is very close to the maximum value of
$\mu$ beyond which
HMRI ceases to exist. Here, the instability is 
extremely sensitive, and
any slight deviation of the experimental conditions from the
numerical assumptions can play a significant role.

For $I_{\rm rod}=4000$ A, the amplitude of the velocity 
is very small
in good accordance with the numerics which predicts stability
for all values of $Re$.

As noted in the introduction, 
the importance of MRI lies in the fact 
that it destabilizes 
rotational flows
with outward increasing angular momentum 
which would otherwise 
be hydrodynamically stable, 
according to
Rayleigh's criterion. 
Doing experiments on MRI it is necessary to validate
that the angular momentum of the base flow
is indeed increasing outward.
This is all the more important since 
a peculiarity of HMRI at small Reynolds and Hartmann numbers 
is that it works only for values of $\mu$ which are
slightly above the Rayleigh line.
Since, in a finite system, some deviation from the
ideal TC-flow profile is unavoidable, 
in particular close to the end-caps, one has to be
careful that the observed instability is indeed 
a magnetically triggered one and
not an instability of a flow which is (at least in some
parts) hydrodynamically unstable.

Unfortunately, the UDV technique allows us only 
to measure axial velocity 
distributions in detail. Some efforts have been made to
measure $V_{\varphi}$ by means of
an electric potential probe inserted into the liquid metal, 
but neither the necessary accuracy nor the desired spatial
resolution could  be achieved by this method. For that 
reason we 
have to rely on the numerical 
simulations exclusively. Given the good correspondence of 
numerics and experiment, as seen so far in the 
$V_z$ data (at least for not too high values of $Re$), 
we believe that the numerics
gives a more or less true picture
of the real physical process.

In the following we will consider the particular 
case  $\mu=0.27$, 
$I_{\rm coil}=76$ A, $I_{\rm rod}=7000$ A with two 
different values of $f_{\rm in}=0.04$ Hz  (corresponding 
to $Re=1183$),
and $f_{\rm in}=0.05$ Hz  (corresponding 
to $Re=1479$).
For these two cases we plot 
in Fig.\ \ref{Re_angular_momentum}
the numerical snapshots of the 
angular velocity $\Omega(r,z)$, of the specific
angular momentum $r^2 \Omega(r,z)$, and of the radial derivative of 
this specific angular momentum.

While the plots for $\Omega(r,z)$ (Fig.\ \ref{Re_angular_momentum}a) 
are not very instructive and 
seem to indicate only a slight difference between the two cases,
the plots for the angular momentum and its radial derivative 
show considerably more structure.
For $f_{\rm in}=0.04$ Hz we observe a rather regular 
outward increase of 
the angular momentum, with the only exception 
close to the splitting positions of the 
upper and lower end-caps. This is not surprising since, with the 
inner ring rotating with the (fast) inner cylinder and the outer ring
rotating with the (slow) outer cylinder, there must be a certain
region where the angular momentum decreases, while in the rest
of the fluid it increases outward.

The crucial point here is the transition from global stability to
global instability when going from  $f_{\rm in}=0.04$ Hz
to  $f_{\rm in}=0.05$ Hz. For the latter we observe 
a wave that travels throughout the total volume of the TC cell.
Clearly visible is the ``fingering'' of the 
angular momentum which is typical for MRI \cite{KNOBLOCHJULIEN}.

\subsection{Variation of $\mu$}

Without any magnetic field, the case $\mu=0$ 
corresponds to the classical 
TC flow with its  well-known critical 
Reynolds number $Re_{\rm crit}=68.2$ (for our 
particular case 
$r_{\rm in}/r_{\rm out}=0.5$)
\cite{CHANDRA2}. With a purely axial or a purely 
azimuthal field applied this
critical value would increase. If both field are 
applied, the character of the
instability changes from a steady TVF to a 
travelling wave, although with a very small frequency.
If $\mu$ is then increased, the wave frequency increases also.
HMRI means that the travelling 
wave continues into the region beyond the 
Rayleigh point $\mu_{\rm Rayl}:=r^2_{\rm in}/r^2_{\rm out}=0.25$ 
where any purely 
hydrodynamic instability would
disappear according to Rayleigh's criterion.

For the PROMISE 1 experiment the behaviour of the 
wave with increasing $\mu$ 
was documented in Fig.\ 5 of \cite{NJP}. It was characterized by 
a rather broad range 
of instability extending to values as large as $\mu=0.35$.
Here we present the corresponding results 
for the PROMISE 2 experiment.
Figure \ref{mulinie} shows 12 panels for
$0<\mu<0.35$. We have fixed $f_{\rm in}=0.1$ Hz, 
$I_{\rm coil}=76$ (i.e. $Ha=12$). 
The current through the central rod has been set to
$I_{\rm rod}=7000$ A (i.e. $\beta=4.53$).
It is clearly visible that the
travelling wave continues beyond the Rayleigh line, though 
not very far. At $\mu=0.32$ it has disappeared completely.

The comparison 
of the upward travelling wave  for $\mu=0.26$ and $\mu=0.27$
is illustrated in Fig.\ \ref{muvergleich}.
Figures \ref{muanalyse} and \ref{muanalyse_detail}
show again the squared rms of the velocity, the squared ratio of 
the upward and 
downward travelling wave,
the wave frequency and the wave speed. In addition to the results of a
time-stepping 2D solver we have also added the critical $\mu$ values for the 
convective and the absolute instability as they result
from a 1D eigenvalue solver \cite{PRIEDE2}. 

Besides the good agreement of experimental and 
numerical results from the 2D code,
it is most remarkable that the transition 
point from the 2D code  corresponds quite
accurately to the absolute
instability from the 1D simulation, while the 1D results 
for the 
convective instability lie at much too high $\mu$ values.
This is strong evidence that the observed instability 
is a global one and not only a noise-triggered
convective one as claimed in \cite{LIU3}.

\subsection{Variation of $\beta$}

For liquid metals with $Pm=10^{-6}...10^{-5}$, 
the critical Reynolds number for 
SMRI 
(with a purely axial magnetic field)
is in the order of several million. The key point of HMRI
is that by
adding an azimuthal magnetic field, this number drops 
drastically to approximately $10^3$.
For PROMISE 1, the transition to instability for 
increasing $\beta$ turned out to be not very sharp, as 
documented in \cite{APJL}. In the following we will
investigate the 
corresponding 
transition for PROMISE 2.

Figure \ref{betalinie} shows 12 panels for
$0<I_{\rm rod}<8200$ A. 
Evidently, there is no indication of any instability 
for $I_{\rm rod}<3000$ A. 
The transition to MRI occurs approximately 
between $I_{\rm rod}=4000$ and $I_{\rm rod}=5000$.
The comparison with the numerical simulation                   %!!! FFT weg
for $I_{\rm rod}=4500$ A and $I_{\rm rod}=7000$ A is shown
in Fig. \ref{betavergleich}.

Figure \ref{betaanalyse} 
shows again the quantitative analysis for two different 
lines corresponding to $\mu=0.26$ and $\mu=0.27$.
The first noticeable fact is that for $\mu=0.26$ we obtain, 
quite similar to the case of the $\mu$ line discussed above, 
a good agreement of the 1D numerical 
results for the absolute instability and the 
2D simulation, while the 1D code results for the convective 
instability would point to much smaller values of the critical 
$\beta$.
Quite  interesting, however, is the fact that for $\mu=0.27$ the
agreement of absolute instability (at $I_{\rm rod} \sim 8200$ A) 
and global instability 
($I_{\rm rod} \sim 7000$ A) becomes worse.
Evidently, the value $\mu=0.27$ is 
(at least for the given $Ha$ number) very close to the 
upper limit of the HMRI. In this case, the threshold of MRI 
will be extremely sensitive to any slight modification of the 
base $V_{\varphi}(r)$ profile. This is a typical case in which we 
indeed observe a global instability for the experimental 
axial boundary conditions, but the hypothetical 
use of ideal Taylor-Couette boundary conditions 
would make the flow stable again. 
This interesting case has been observed in the 
numerical simulations of Liu (\cite{LIU2} for PROMISE 1,
\cite{LIU3} for PROMISE 2).
However, as we see here the interpretation of the author is 
not correct: what is observed for real boundary conditions 
is not a 
noise-induced convective instability (this would start already 
at $\beta=2.3$) but still a
global instability, although on the
basis of a flow that is slightly modified by the 
lids which make the flow a bit more prone to MRI
than the ideal TC-flow would be.

\subsection {Variation of $Ha$}

MRI is
a weak field instability characterized not only by 
a lower
critical $Ha$ but also by an upper one. 
For the PROMISE 1 
experiment it had been shown in  \cite{PRL}
that the travelling 
wave appears approximately at a coil 
current 30 A and disappears again at 110 A. 

Here we discuss the corresponding behaviour for PROMISE 2. 
We choose $f_{\rm in}=0.06$ Hz, $\mu=0.27$, and study in detail the
cases
$I_{\rm rod}=7000$ A and $I_{\rm rod}=8000$ A.
Figures \ref{Halinie}, \ref{Havergleich}, and 
\ref{Haanalyse} document again the MRI dependence on $I_{\rm coil}$ 
(i.e. $Ha$), the comparison with numerics, and                       %!!!
the quantitative analysis, respectively.
While the transition between stable and unstable regime
for large $Ha$ numbers (around 15) shows a 
satisfactory agreement of experiment
and numerics, this agreement becomes worse for the transition at
low $Ha$ numbers. Actually, the observed MRI
starts already at lower $Ha$ (around 4...5) in contrast to the theory 
which predicts it to start between 8 and 9.
This effect needs further work to be understood.

Similar as for the $\beta$-line (at $\mu=0.27$) 
in the former subsection, we have to note that
the absolute instability would only start slightly beyond 
the chosen rod-current of 8000 A. We obtain, nevertheless a global 
instability which points to a slight modification of the base flow.

\section{Conclusions}

In this paper we have documented 
and analyzed four types of
parameter variations in order to characterize in detail the 
HMRI in a TC-flow with split end-caps (PROMISE 2).
Due to the strongly reduced Ekman pumping,  
resulting from this apparently minor modification,
we have observed much sharper transitions from 
the stable flow regime to the MRI regime  
than in the former PROMISE 1 experiment. 

In general, the experimentally 
observed thresholds turned out to be
in good correspondence 
with the thresholds resulting from  
2D simulations for the global instability.
In addition to that, we have shown that the 
results of a  1D eigenvalue solver for the
absolute instability are in good agreement with the 
global instability, 
at least as long as we are not too close 
to some upper limit values of $\mu$ 
(approximately ~0.27 for the
considered values of $Re$, $Ha$ and $\beta$)
for the occurrence of HMRI.

In contrast to that, the convective instability resulting from
the 1D code would generally imply a larger range of the 
instability
than observed in the experiment.
From this it seems justified to 
conclude that the observed instability is indeed 
a global one which corresponds, in a wide range of parameters, 
to the absolute instability and not to a noise-driven 
convective one.

Coming closer to those parameters (in particular $\mu$) for 
which helical MRI ceases to exist (at least for the 
achievable values of $Ha$ and $\beta$) we observe increasing 
deviations between global and absolute instability, which 
very likely trace back to the slightly changed velocity 
profile of the base flow in the finite TC-cell.

 What is left for future work is a careful 
comparison of the $z$-dependence of the wave 
amplitude with the corresponding results of the 2D 
code and of the 1D eigenvalue code (in which
this z-dependence appears as an imaginary 
part of the wavenumber). Comparisons for a few
parameters seem to indicate a fairly good agreement, 
but a quantitative comparison for all parameter 
variations requires much more work.

A ``promising'' perspective of PROMISE would be 
to extend  the range of 
parameters $Re$ and $Ha$ 
in order to 
investigate  the continuous transition between 
the well-studied region of 
HMRI and the experimentally much more demanding 
region of SMRI. One way to accomplish this would be to 
replace  the working fluid GaInSn by sodium.
Another interesting extension of the work would be to
study various 
combinations of MRI and Tayler instability, 
which would require to direct electric 
currents through the liquid metal.
The most interesting idea, however, namely 
to realize in experiment 
a  self-sustaining nonlinear dynamo process 
in an MRI-destabilized (quasi)-Keplerian 
shear flow \cite{RINCON}, will probably remain a 
dream, at least for liquid metal flows.
Some appropriate plasma experiments \cite{FOREST} 
may be better suited
for such an ambitious project.

\acknowledgments

This work was supported by the German 
Leibniz Gemeinschaft, within its Senatsausschuss 
Wettbewerb (SAW)
programme, and by the Deutsche Forschungsgemeinschaft in the framework of 
SFB 609.

\pagebreak

\begin{figure}
\begin{center}
\epsfxsize=8.6cm\epsfbox{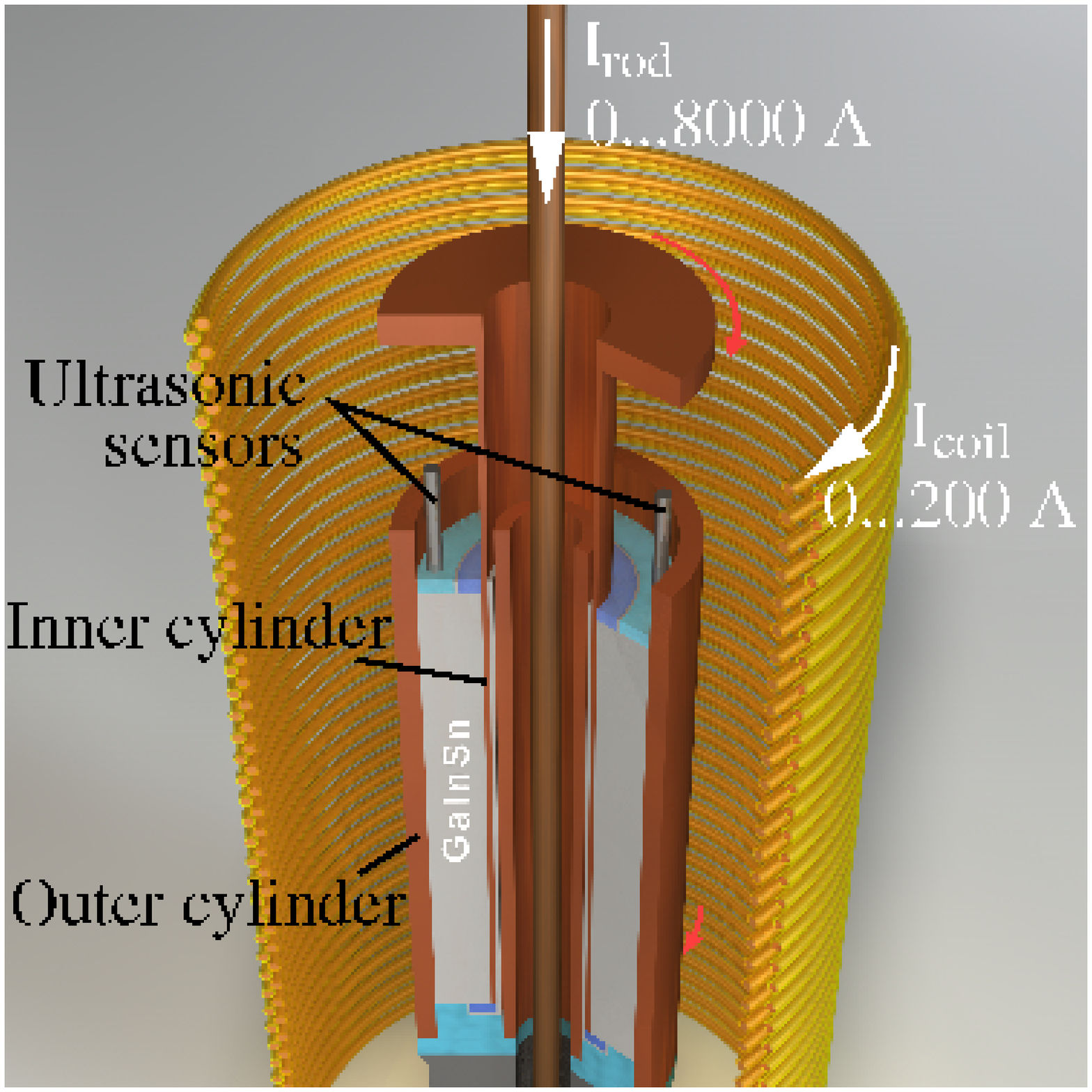}\\
\vspace{2mm}
\caption{Sketch of the PROMISE 2 experiment. 
The gap width  of the Taylor-Couette cell is 40 mm, 
the height is 400 mm.}
\label{set-up1}
\end{center}
\end{figure}

\begin{figure}
\begin{center}
\epsfxsize=8.6cm\epsfbox{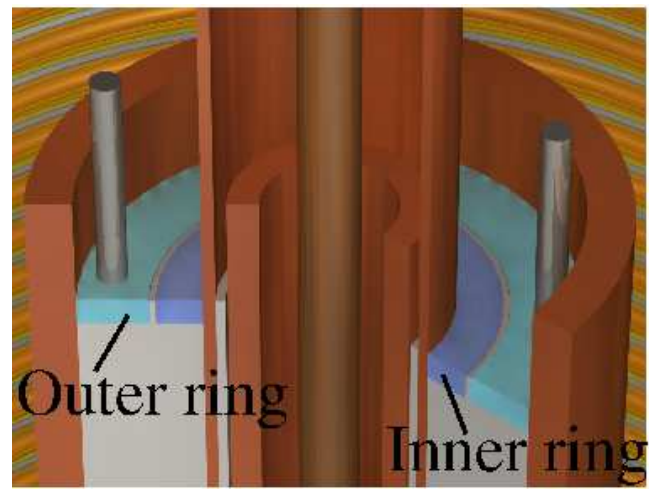}\\
\vspace{2mm}
\caption{Details of the PROMISE 2 experiment at the top. 
The inner ring rotates
with the inner cylinder, the outer ring rotates with the 
outer cylinder. The same applies to the end-cap at the bottom.}
\label{set-up2}
\end{center}
\end{figure}

\clearpage
\pagebreak

\begin{figure}
\begin{center}
\epsfxsize=8.6cm\epsfbox{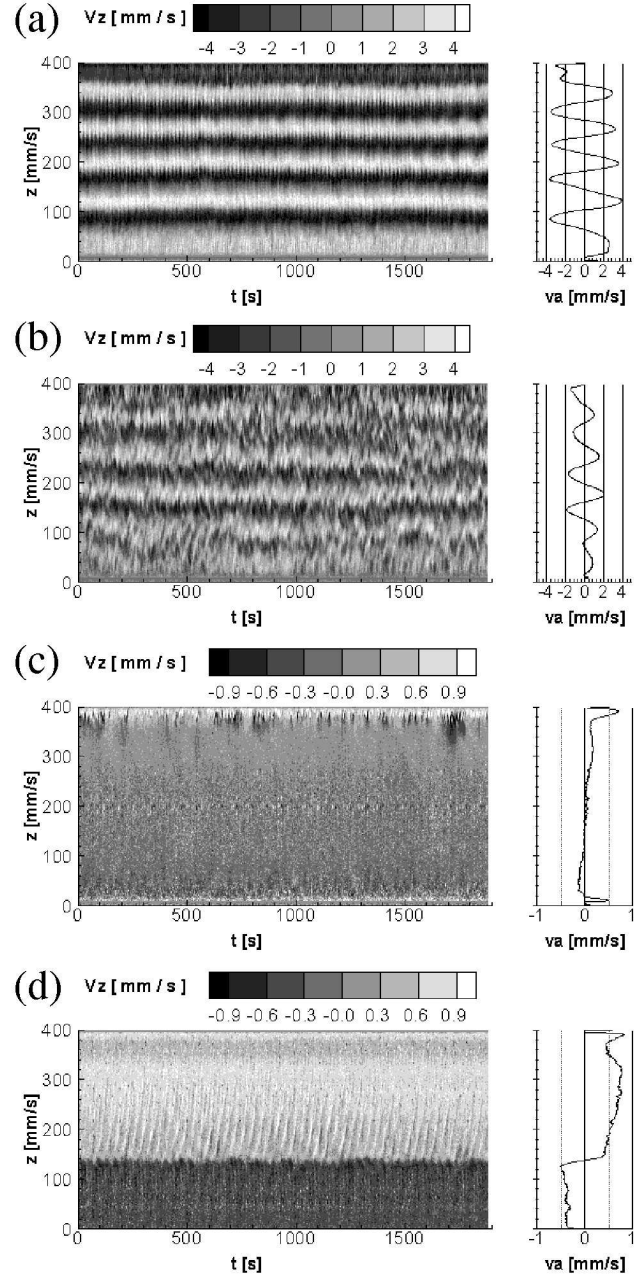}
\vspace{2mm}
\caption{Measured axial velocity $V_z(z,t)$ (l.h.s.), and 
time averaged $va_z(z)$ (r.h.s.),
in the field-free case for PROMISE 2, 
at $f_{\rm in}=0.1$ Hz: (a) $\mu=0$, (b) $\mu=0.1$, 
(c) $\mu=0.27$. (d) The same for 
PROMISE 1 at $f_{\rm in}=0.06$ Hz and $\mu=0.27$.}
\label{withoutfield}
\end{center}
\end{figure}

\clearpage
\pagebreak

\begin{figure}
\begin{center}
\epsfxsize=8.6cm\epsfbox{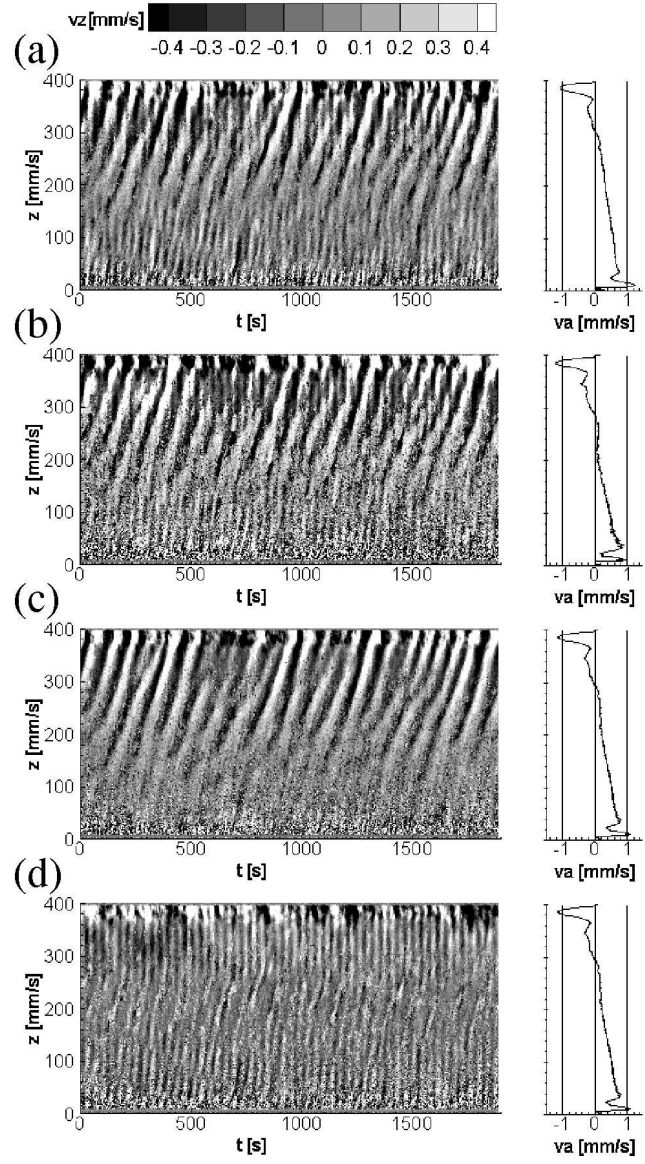}
\vspace{2mm}
\caption{Measured axial velocity for 
$f_{\rm in}=0.1$ Hz, $f_{\rm out}=0.027$ Hz, 
$I_{\rm rod}=7000$ A, and 
$I_{\rm coil}=76$ A.
(a) at UDV sensor 1, (b) at UDV sensor 2, (c) average of 
sensor 1 and sensor 2, (d) difference of sensor 1 and sensor 2. 
The frequency seen in (d) is just the rotation rate of the 
outer cylinder.}
\label{individualsensors}
\end{center}
\end{figure}

\clearpage
\pagebreak

\begin{figure*}
\begin{center}
\epsfxsize=17.2cm\epsfbox{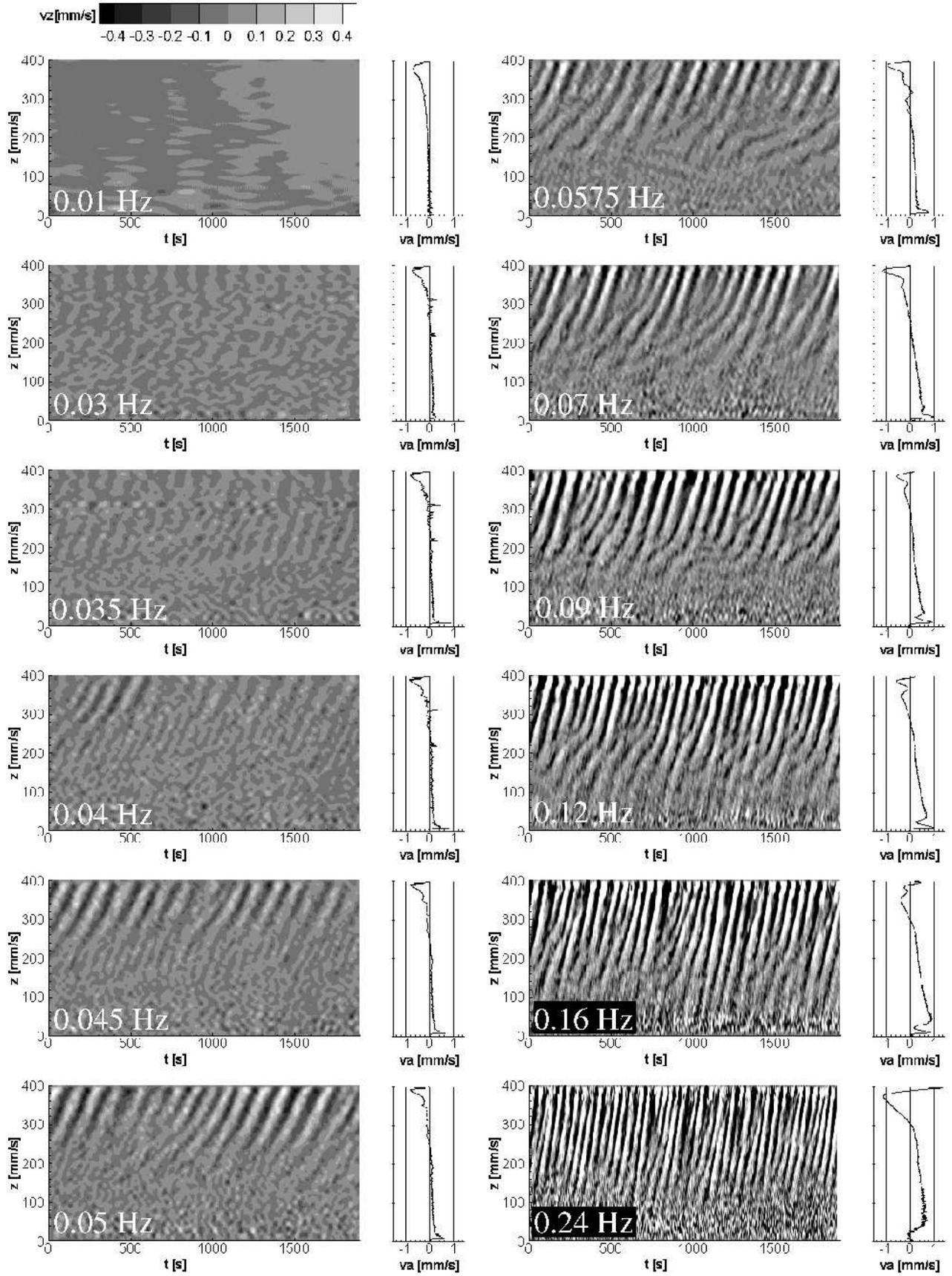}
\vspace{2mm}
\caption{$v_z(z,t)$ and $va_z(z)$, both 
averaged over the two UDV sensors, 
in dependence on $f_{\rm in}$ (i.e. $Re$) 
for $\mu=0.27$, $I_{\rm rod}=7000$ A, and 
$I_{\rm coil}=76$ A.}
\label{Relinie}
\end{center}
\end{figure*}

\clearpage
\pagebreak

\begin{figure*}
\begin{center}
\epsfxsize=8.6cm\epsfbox{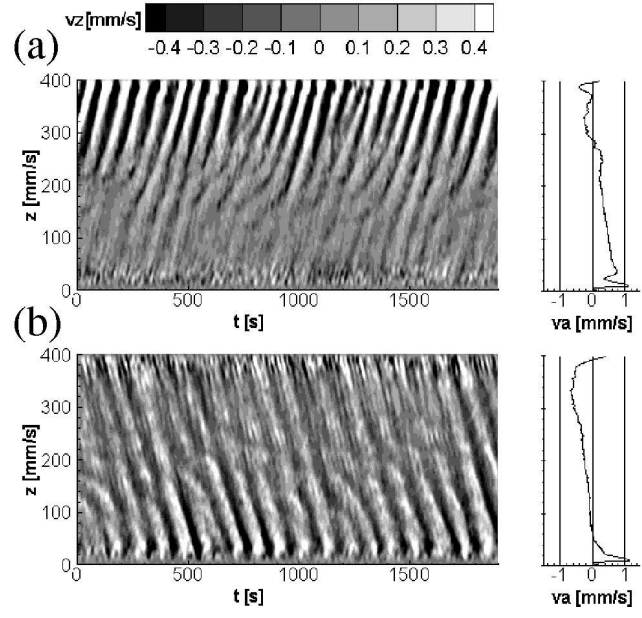}
\vspace{2mm}
\caption{$v_z(z,t)$ and $va_z(z)$, both 
averaged over the two UDV sensors, 
for $f_{\rm in}=0.1$ Hz,
$\mu=0.27$, $I_{\rm rod}=7000$ A: (a) $I_{\rm coil}=76$ A, 
(b) $I_{\rm coil}=-76$ A.
The change of the current direction in the coil makes the 
MRI wave change its direction from (a) upward to (b) downward.}
\label{direction}
\end{center}
\end{figure*}

\clearpage
\pagebreak

\begin{figure}
\begin{center}
\epsfxsize=8.6cm\epsfbox{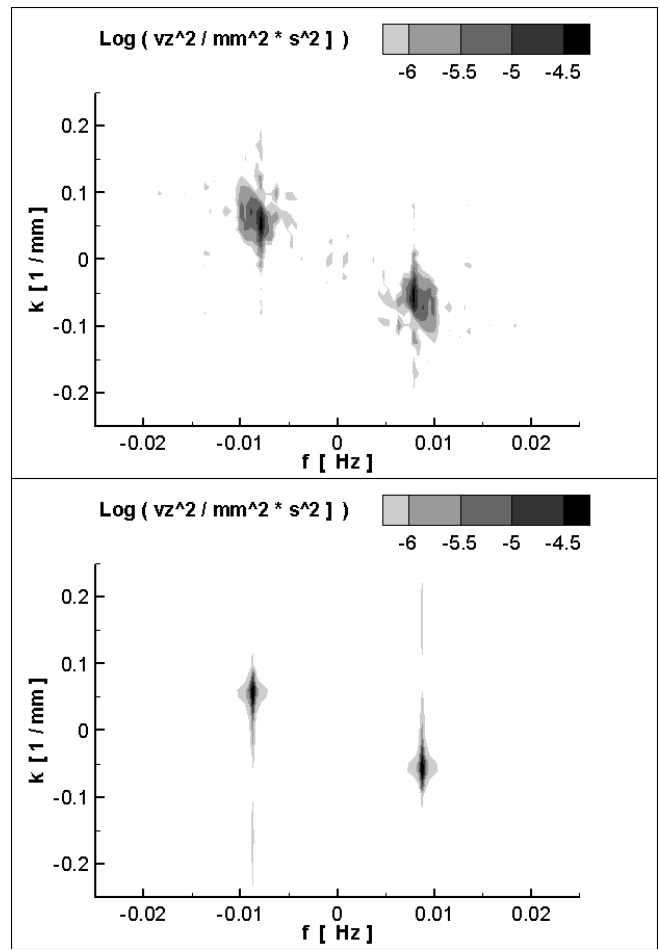}
\vspace{2mm}
\caption{Power spectral density of the 2D FFT of $v_z(z,t)$
for $f_{\rm in}=0.05$ Hz, $\mu=0.27$, $I_{\rm rod}=7000$ A, and 
$I_{\rm coil}=76$ A. Upper panel: experimental results, Lower panel: 
results of the simulation with a 2D solver.}
\label{ReFFT}
\end{center}
\end{figure}

\clearpage
\pagebreak

\begin{figure*}
\begin{center}
\epsfxsize=8.6cm\epsfbox{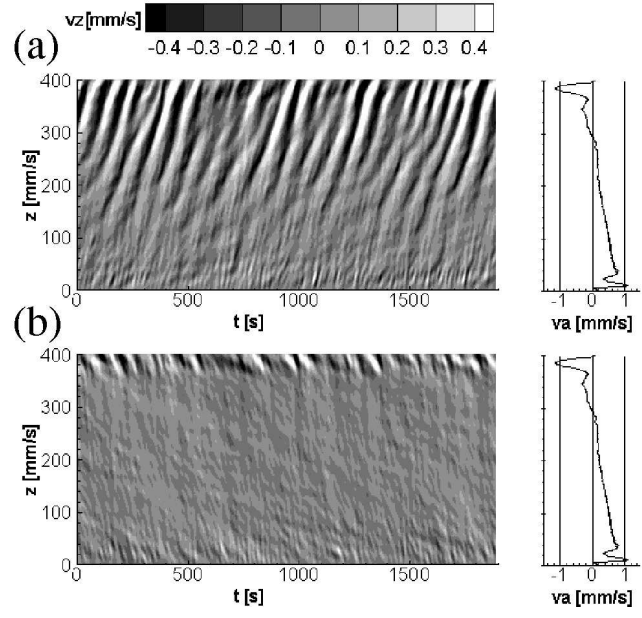}
\vspace{2mm}
\caption{$va_z(z)$ and Hilbert transforms of $v_z(z,t)$ 
for $f_{\rm in}=0.1$ Hz,
$\mu=0.27$, $I_{\rm rod}=7000$ A, $I_{\rm coil}=76$ A. 
(a) Dominant upward travelling component, (b) minor downward travelling
component which results from a reflection of the upward travelling component
at the upper end-cap.}
\label{updown}
\end{center}
\end{figure*}

\begin{figure}
\begin{center}
\epsfxsize=8.6cm\epsfbox{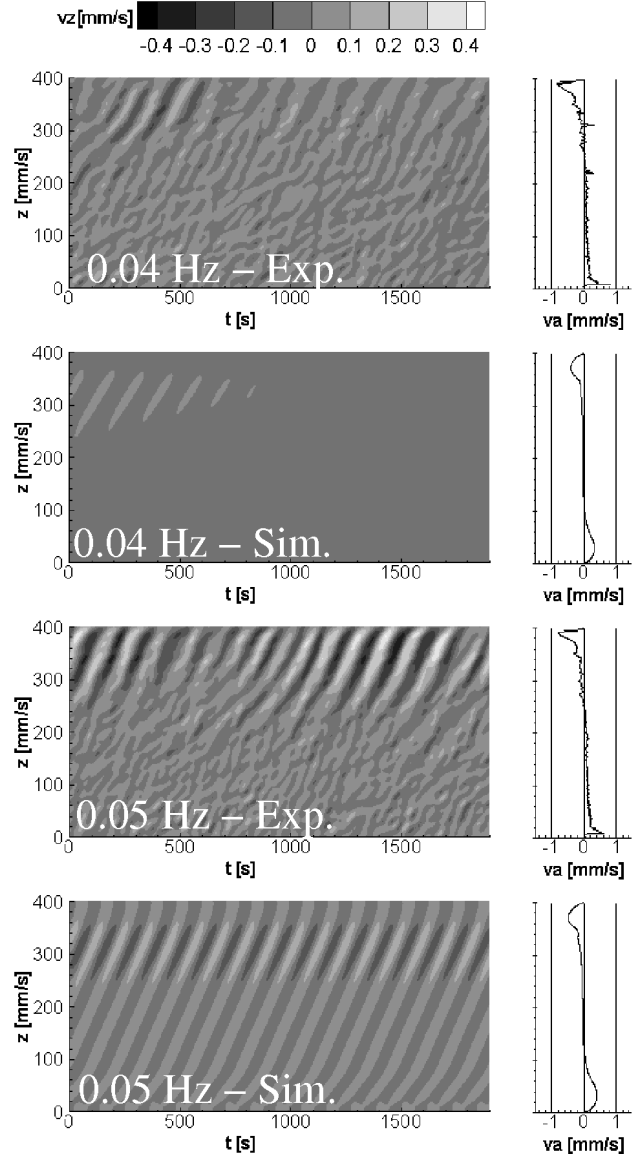}
\vspace{2mm}
\caption{Comparison of experimental and simulated results for the 
upward travelling wave component for $\mu=0.27$, $I_{\rm rod}=7000$ A, 
$I_{\rm coil}=76$ A, 
$f_{\rm in}=0.04$ Hz (upper two panels) and 
$f_{\rm in}=0.05$ Hz (lower two panels).}
\label{Revergleich}
\end{center}
\end{figure}

\clearpage
\pagebreak

\begin{figure}
\begin{center}
\epsfxsize=8.6cm\epsfbox{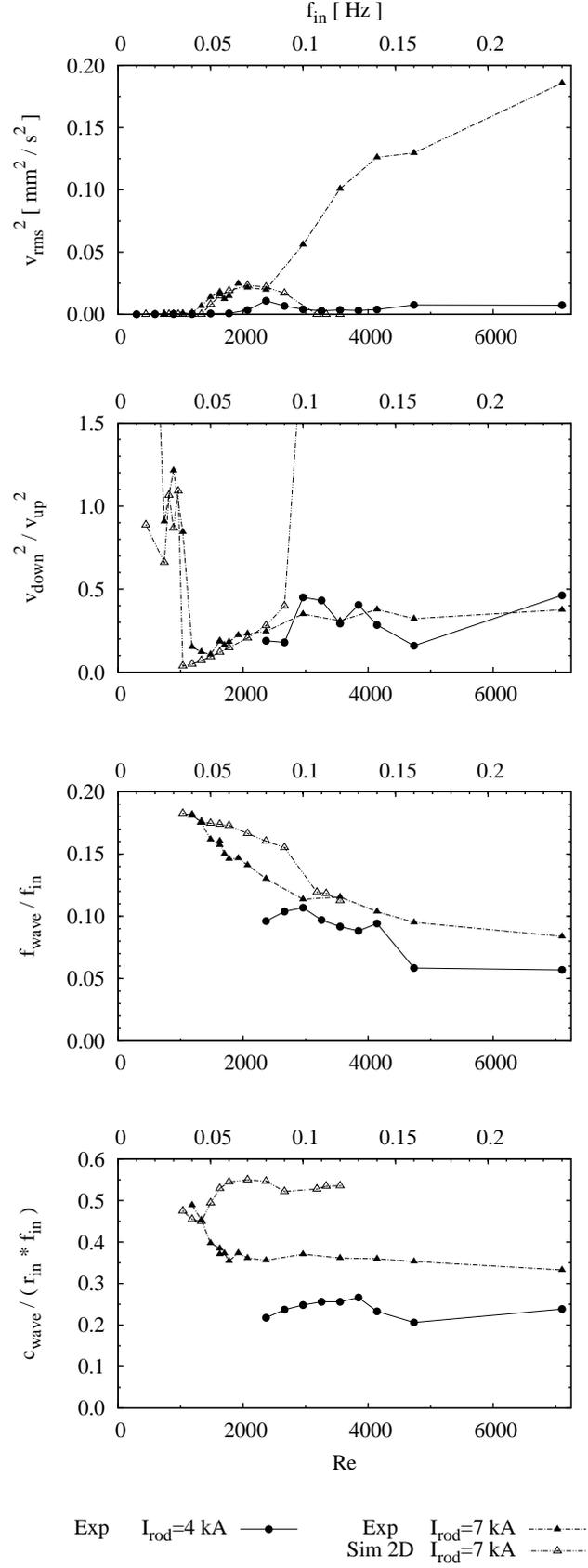}
\vspace{2mm}
\caption{Squared rms of $v_z(z,t)$, squared ratio of
maximum amplitude of downward to upward travelling waves, normalized 
frequency and normalized wave speed for $\mu=0.27$, $I_{\rm rod}=4000$ A 
and 7000 A, 
$I_{\rm coil}=76$ A, in dependence on $f_{\rm in}$ (i.e. $Re$).}
\label{Reanalyse}
\end{center}
\end{figure}

\clearpage
\pagebreak

\begin{figure}
\begin{center}
\epsfxsize=8.6cm\epsfbox{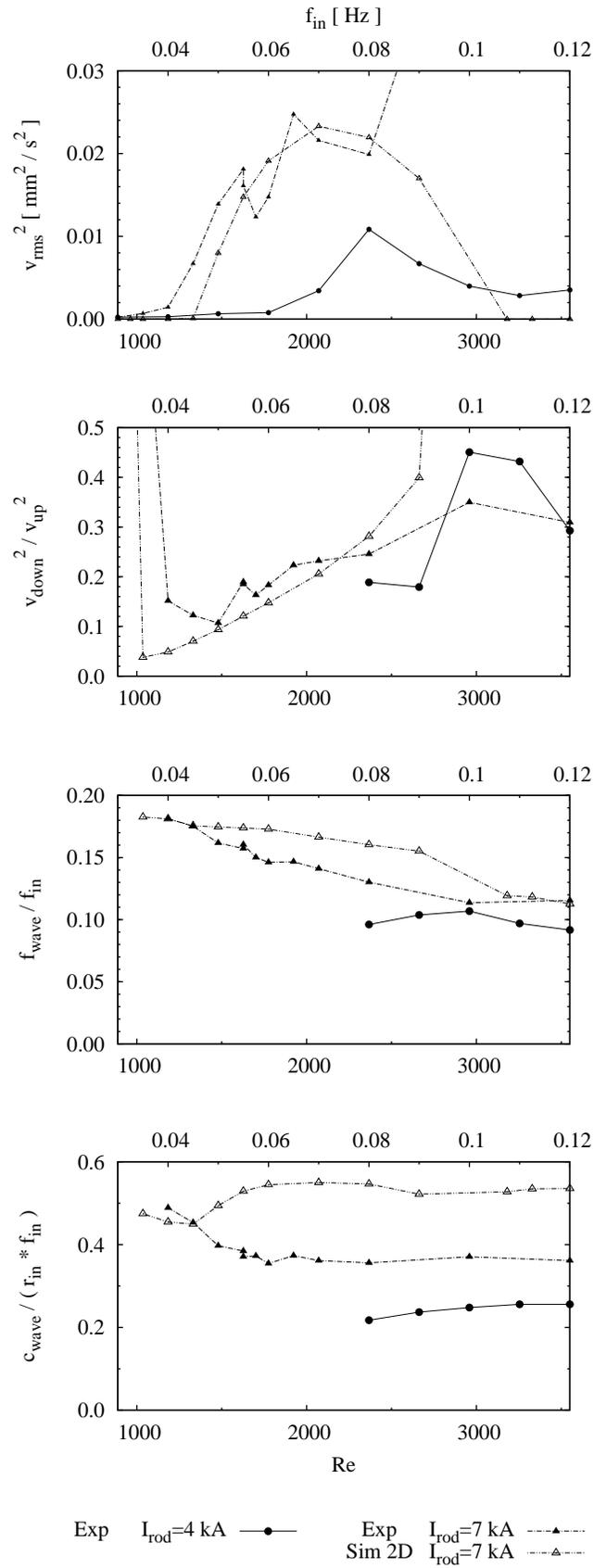}
\vspace{2mm}
\caption{Same as Fig.\ \ref{Reanalyse}, but zoomed to medium values of $f_{\rm in}$.}
\label{Reanalyse_detail}
\end{center}
\end{figure}

\clearpage
\pagebreak

\begin{figure}
\begin{center}
\epsfxsize=8.6cm\epsfbox{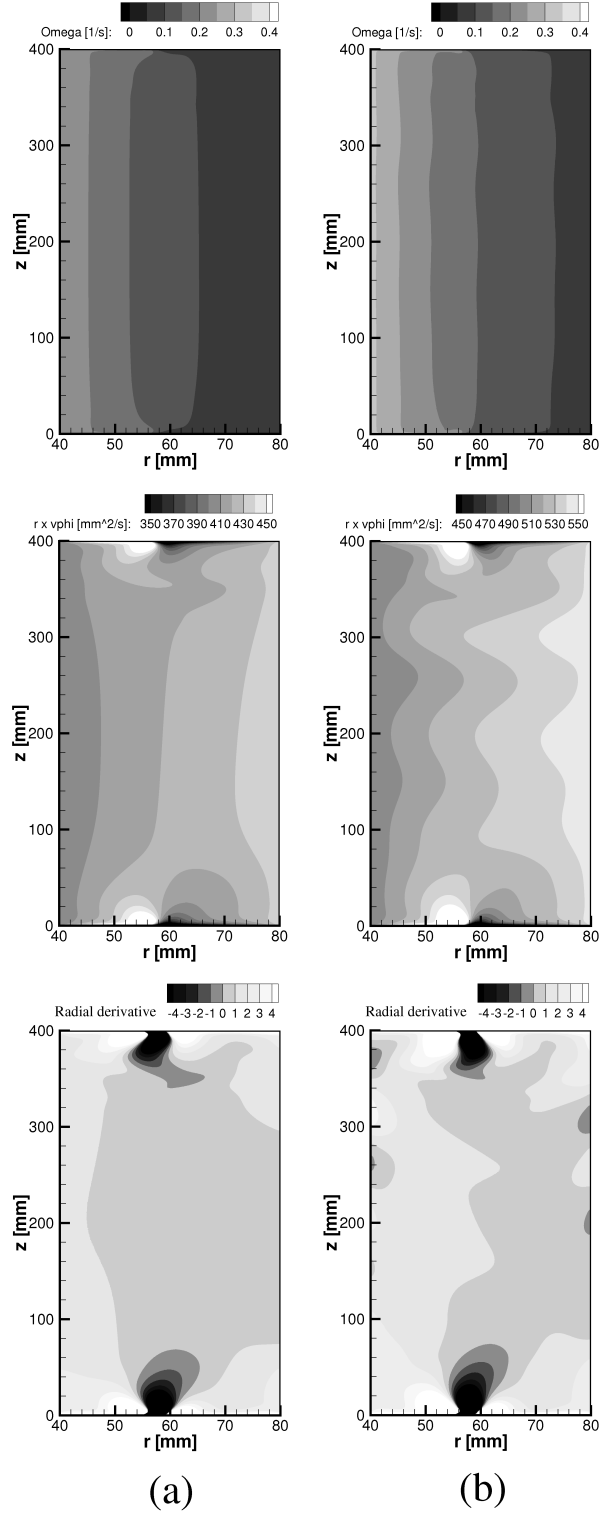}
\vspace{2mm}
\caption{From top to bottom: Simulated angular velocity $\Omega(r,z)$, 
specific angular momentum $r^2 \Omega(r,z)$, and radial derivative of 
$r^2 \Omega(r,z)$ for $\mu=0.27$, $I_{\rm rod}=7000$ A, 
$I_{\rm coil}=76$ A. (a) 
Subcritical $f_{\rm in}=0.04$ Hz. (b) Supercritical 
$f_{\rm in}=0.05$ Hz. }
\label{Re_angular_momentum}
\end{center}
\end{figure}

\clearpage
\pagebreak

\begin{figure*}
\begin{center}
\epsfxsize=17.2cm\epsfbox{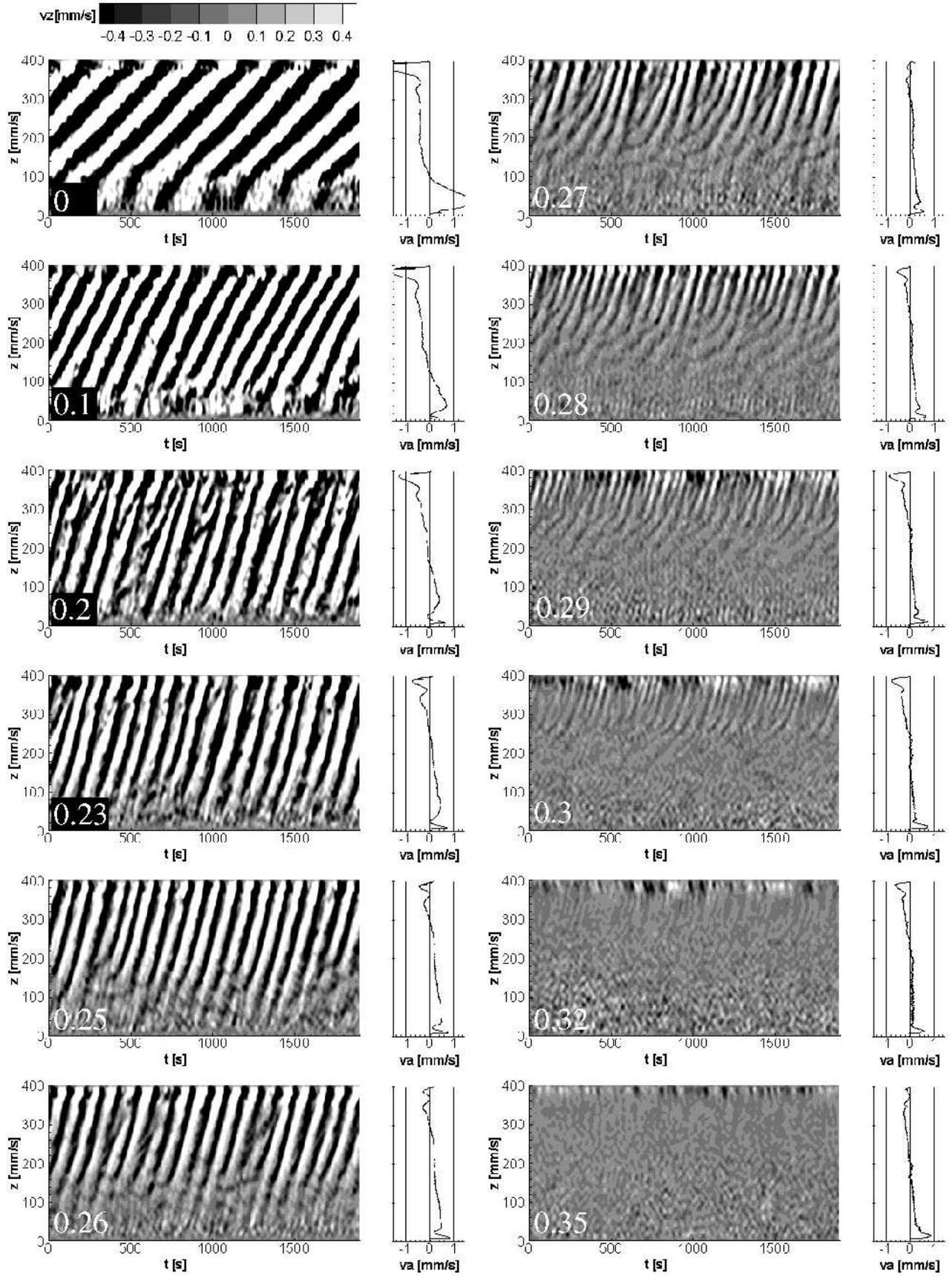}
\vspace{2mm}
\caption{$v_z(z,t)$ and $va_z(z)$, both 
averaged over the two UDV sensors, 
in dependence on $\mu$ for $f_{\rm in}=0.1$ Hz, $I_{\rm rod}=7000$ A, and 
$I_{\rm coil}=76$ A.}
\label{mulinie}
\end{center}
\end{figure*}

\clearpage
\pagebreak

\begin{figure}
\begin{center}
\epsfxsize=8.6cm\epsfbox{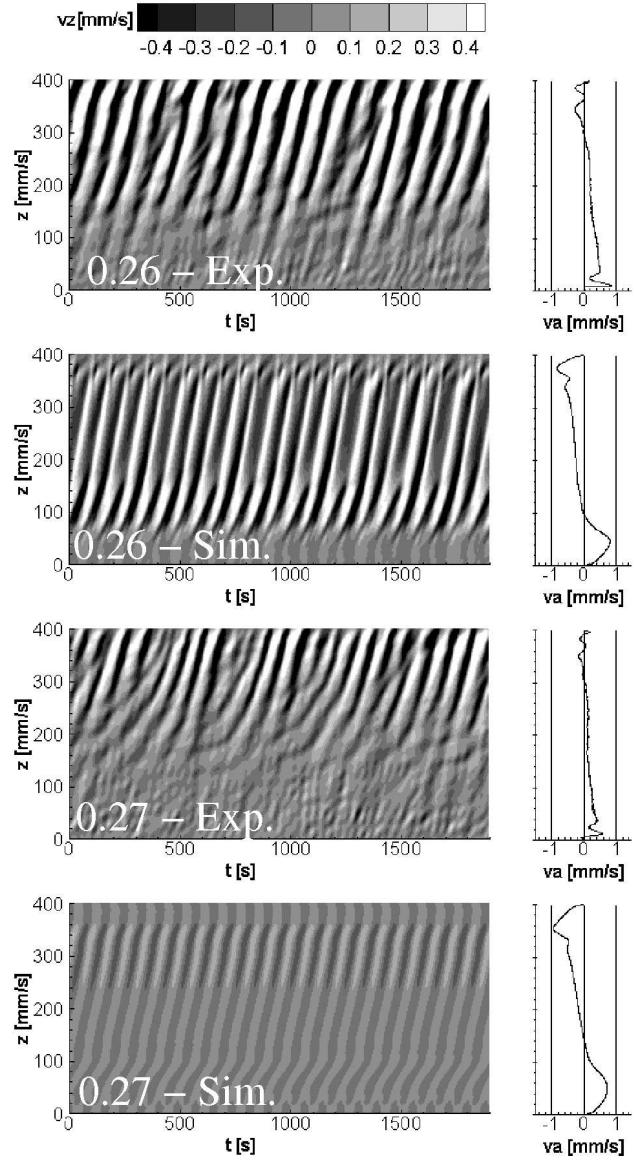}
\vspace{2mm}
\caption{Comparison of experimental and simulated results for the 
upward travelling wave component
for $f_{\rm in}=0.1$ Hz, $I_{\rm rod}=7000$ A, 
$I_{\rm coil}=76$ A, $\mu=0.26$ (upper two panels) and $\mu=0.27$ (lower 
two panels).}
\label{muvergleich}
\end{center}
\end{figure}

\clearpage
\pagebreak

\begin{figure}
\begin{center}
\epsfxsize=8.6cm\epsfbox{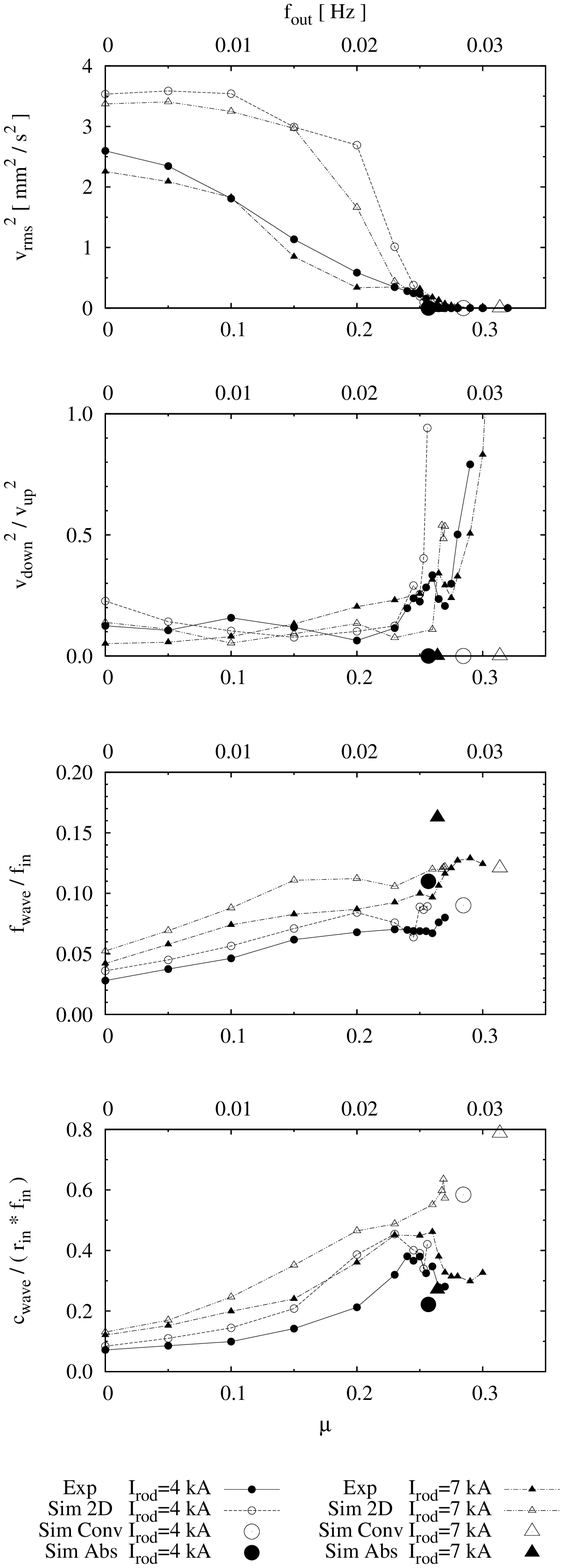}
\vspace{2mm}
\caption{Squared rms of $v_z(z,t)$, squared ratio of
maximum amplitude of downward to upward travelling waves, normalized 
frequency and normalized wave speed for 
$f_{\rm in}=0.1$ Hz, $I_{\rm rod}=4000$ A and 7000 A,  
$I_{\rm coil}=76$ A, in dependence in $\mu$.}
\label{muanalyse}
\end{center}
\end{figure}

\clearpage
\pagebreak

\begin{figure}
\begin{center}
\epsfxsize=8.6cm\epsfbox{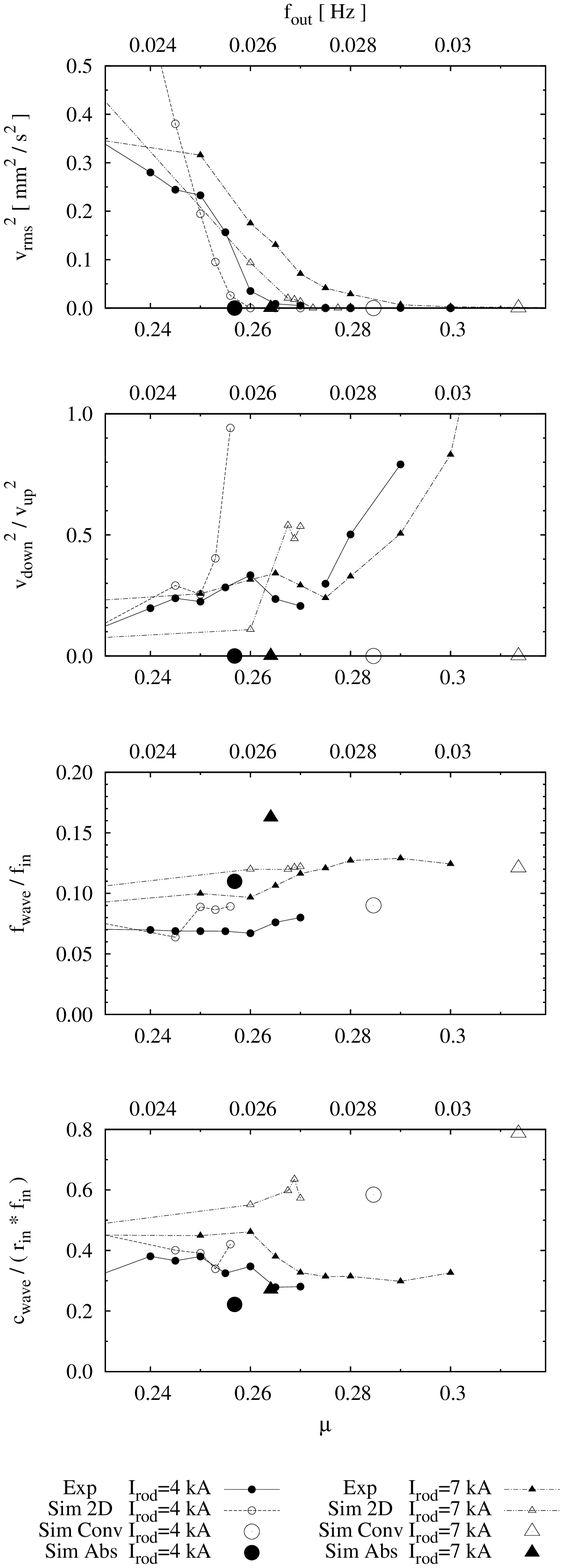}
\vspace{2mm}
\caption{Same as Fig.\ \ref{muanalyse}, but zoomed to $\mu$ values close to 
the Rayleigh point $\mu_{\rm Rayl}=0.25$.}
\label{muanalyse_detail}
\end{center}
\end{figure}

\clearpage
\pagebreak

\begin{figure}
\begin{center}
\epsfxsize=17.2cm\epsfbox{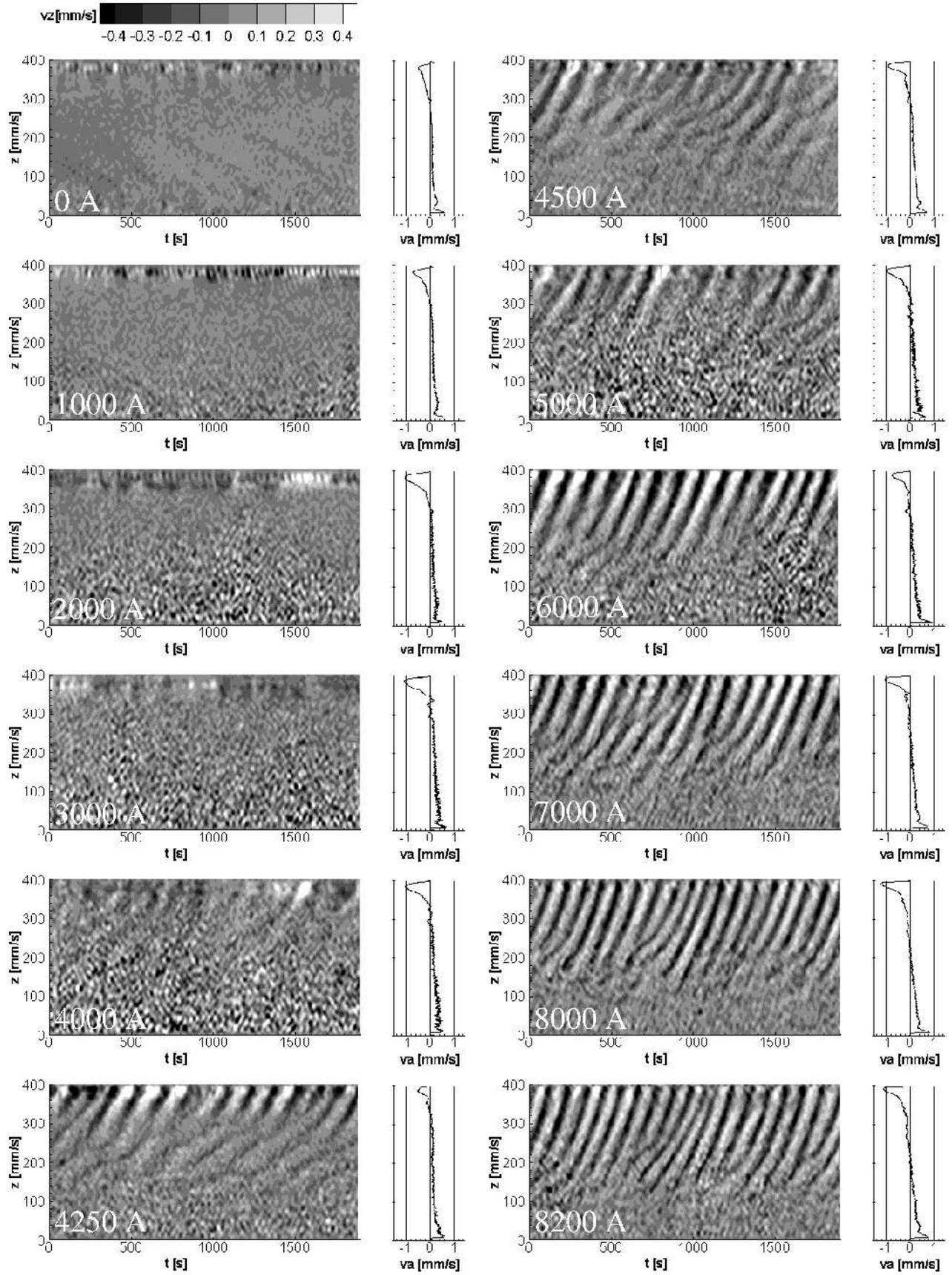}
\vspace{2mm}
\caption{$v_z(z,t)$ and $va_z(z)$, both 
averaged over the two UDV sensors,  
in dependence on $\beta$  for $f_{\rm in}=0.06$ Hz, $\mu=0.26$, and 
$I_{\rm coil}=76$ A.}
\label{betalinie}
\end{center}
\end{figure}
\clearpage
\pagebreak

\begin{figure}
\begin{center}
\epsfxsize=8.2cm\epsfbox{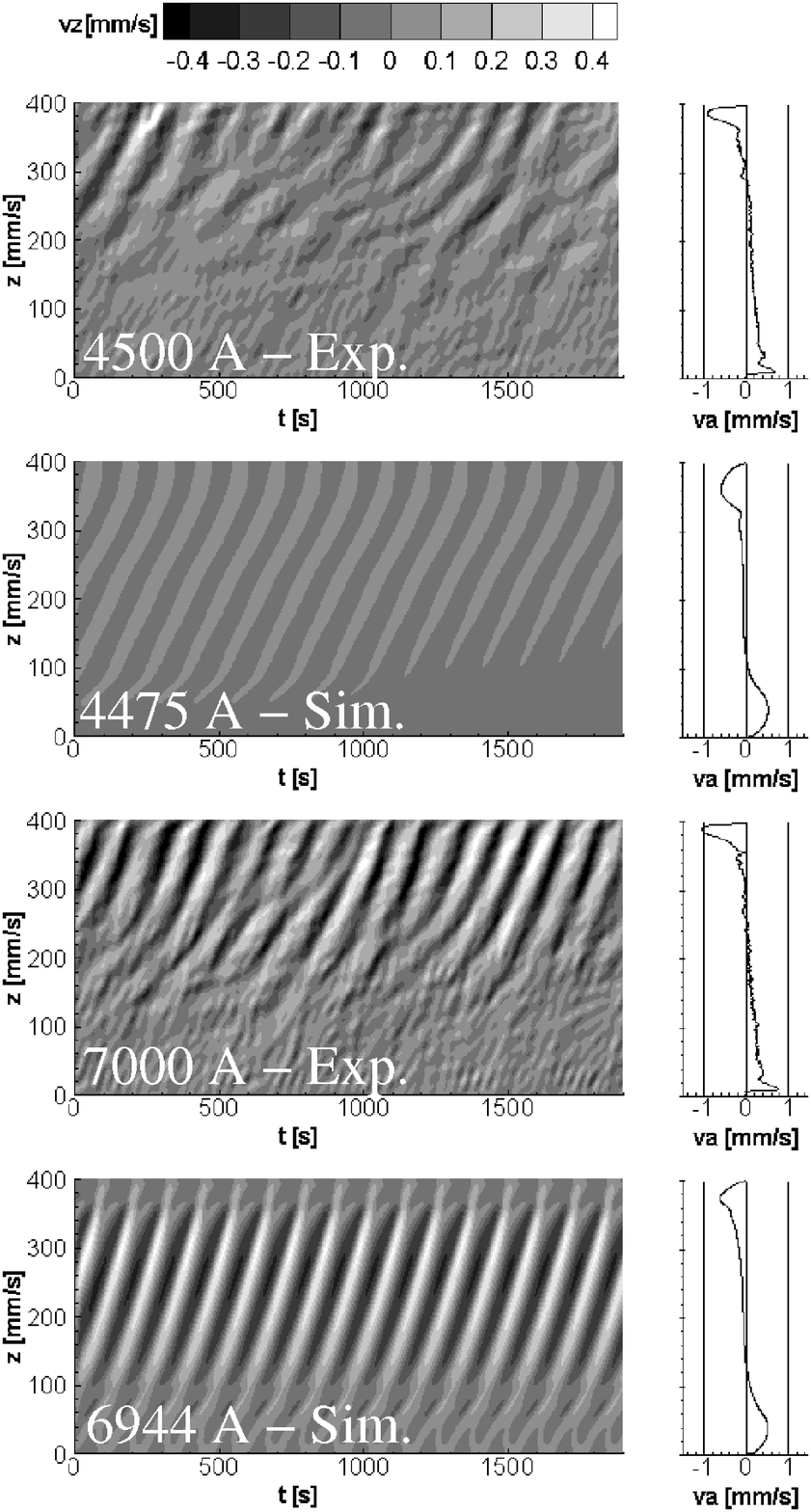}
\vspace{2mm}
\caption{Comparison of experimental and simulated results for the 
upward travelling wave components
for $f_{\rm in}=0.06$ Hz, $\mu=0.26$, and 
$I_{\rm coil}=76$ A,
$I_{\rm rod}\sim 4500$ A (upper two panels) and
$I_{\rm rod}\sim 7000$ A (lower two panels). }
\label{betavergleich}
\end{center}
\end{figure}

\clearpage
\pagebreak
\begin{figure}
\begin{center}
\epsfxsize=8.6cm\epsfbox{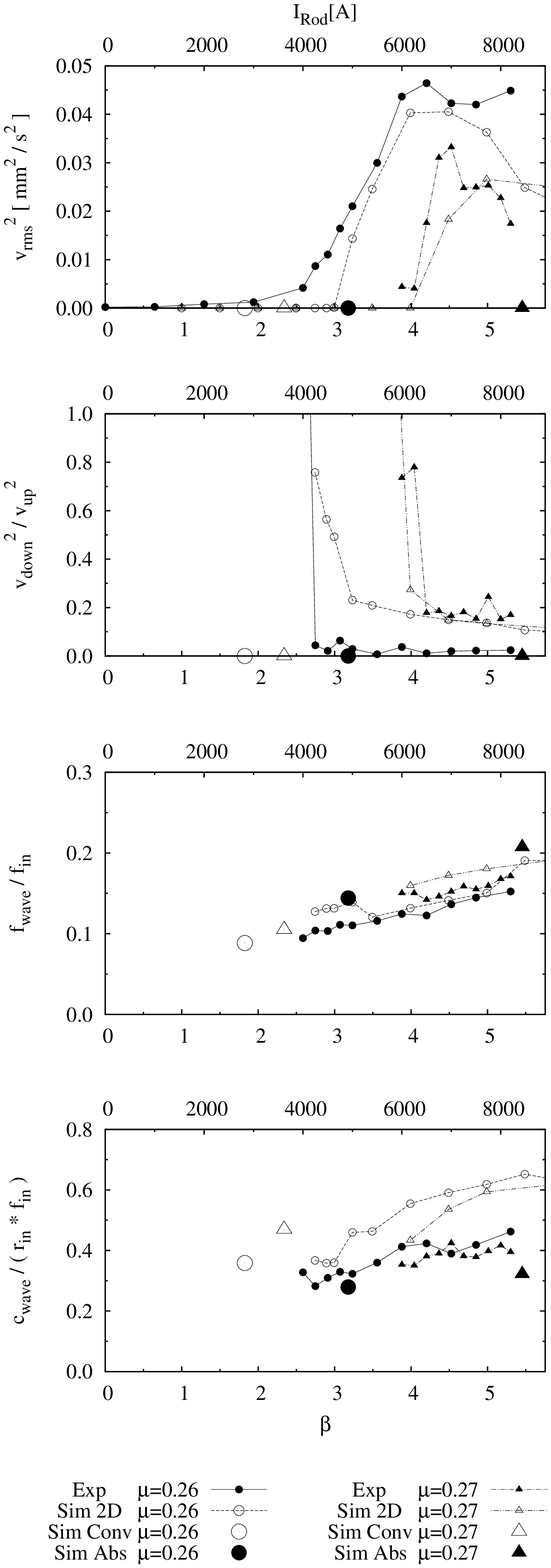}
\vspace{2mm}
\caption{Squared rms of $v_z(z,t)$, squared ratio of
maximum amplitude of downward to upward travelling waves, 
normalized 
frequency and normalized wave speed for
$f_{\rm in}=0.06$ Hz, $\mu=0.26$ and 0.27,
$I_{\rm coil}=76$ A, in dependence on $I_{\rm rod}$ (i.e. $\beta$).}
\label{betaanalyse}
\end{center}
\end{figure}

\clearpage
\pagebreak

\begin{figure*}
\begin{center}
\epsfxsize=17.2cm\epsfbox{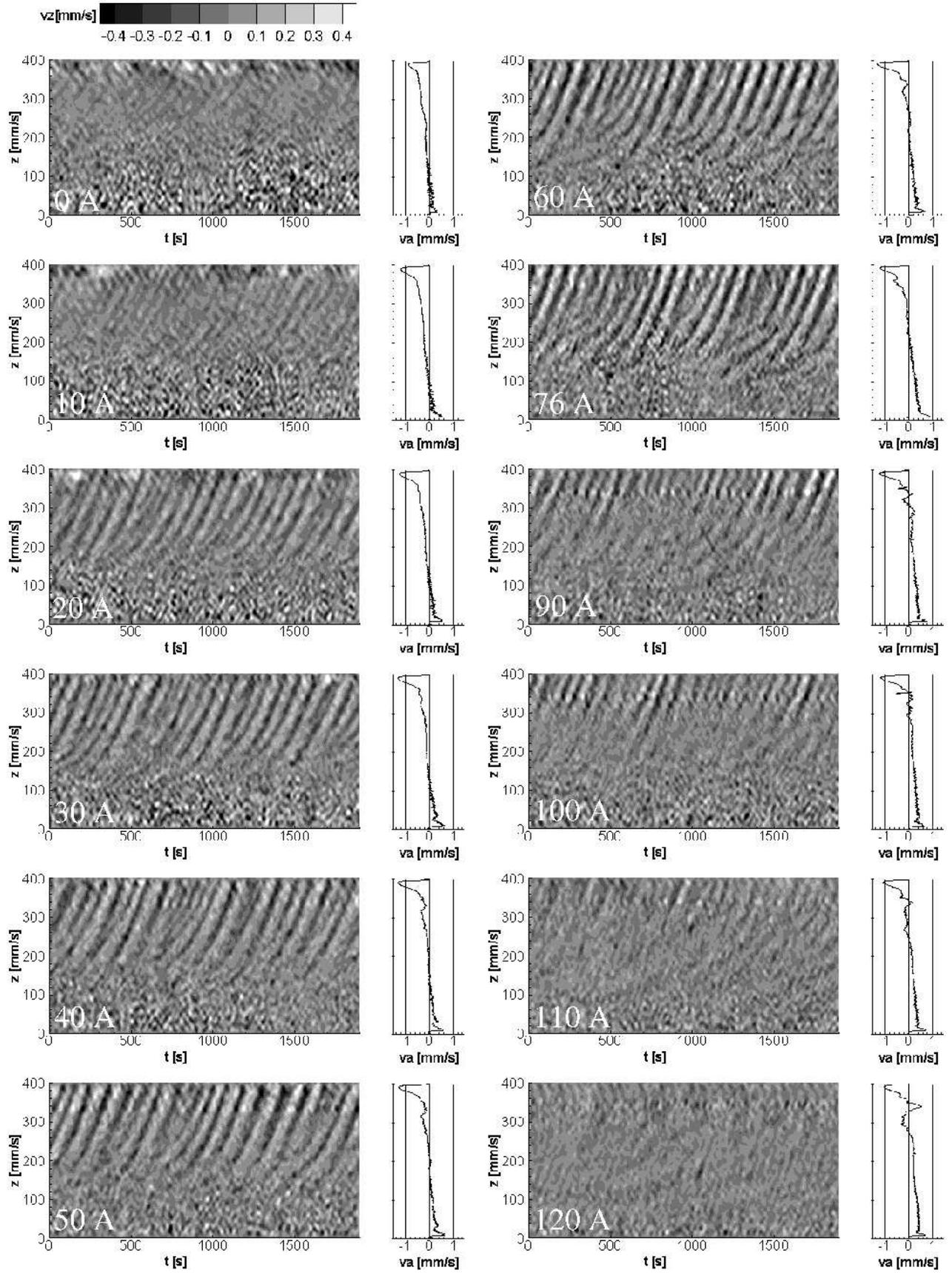}
\vspace{2mm}
\caption{$v_z(z,t)$ in dependence on $I_{\rm coil}$ (i.e. $Ha$) 
 for $f_{\rm in}=0.06$ Hz, $\mu=0.27$, and 
$I_{\rm rod}=8000$ A. }
\label{Halinie}
\end{center}
\end{figure*}

\clearpage
\pagebreak

\begin{figure}
\begin{center}
\epsfxsize=8.6cm\epsfbox{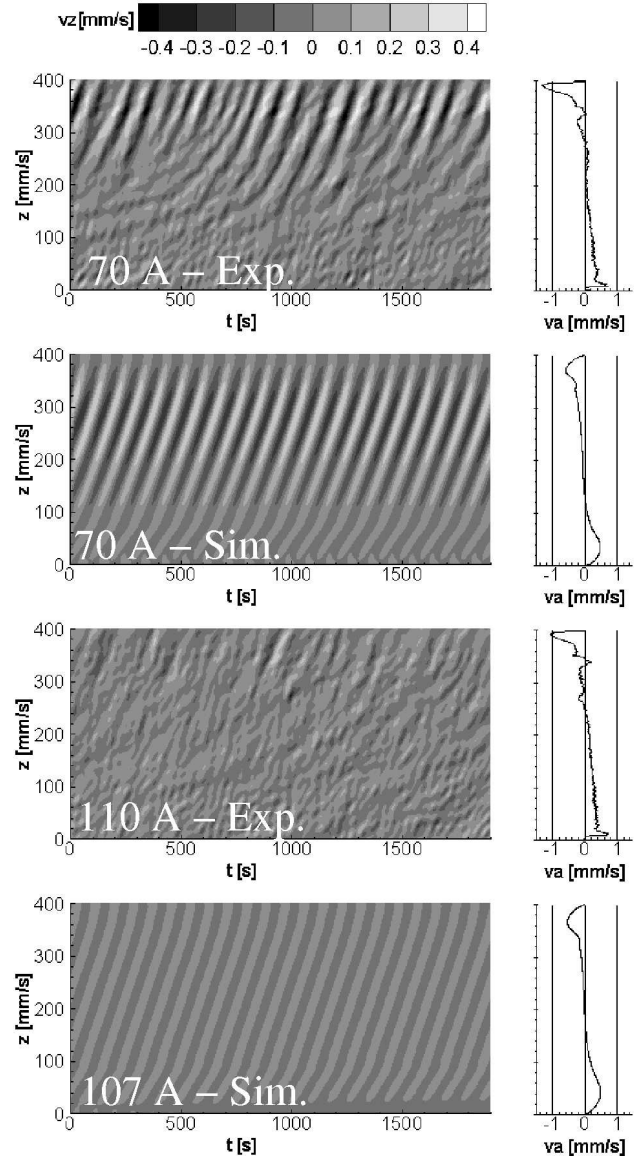}
\vspace{2mm}
\caption{Comparison of experimental and simulated results for the 
upward travelling wave components 
for $f_{\rm in}=0.06$ Hz, $\mu=0.27$, and 
$I_{\rm rod}=8000$ A, $I_{\rm coil}=70$ A and
 $I_{\rm coil}\sim110$ A. }
\label{Havergleich}
\end{center}
\end{figure}

\clearpage
\pagebreak

\begin{figure}
\begin{center}
\epsfxsize=8.6cm\epsfbox{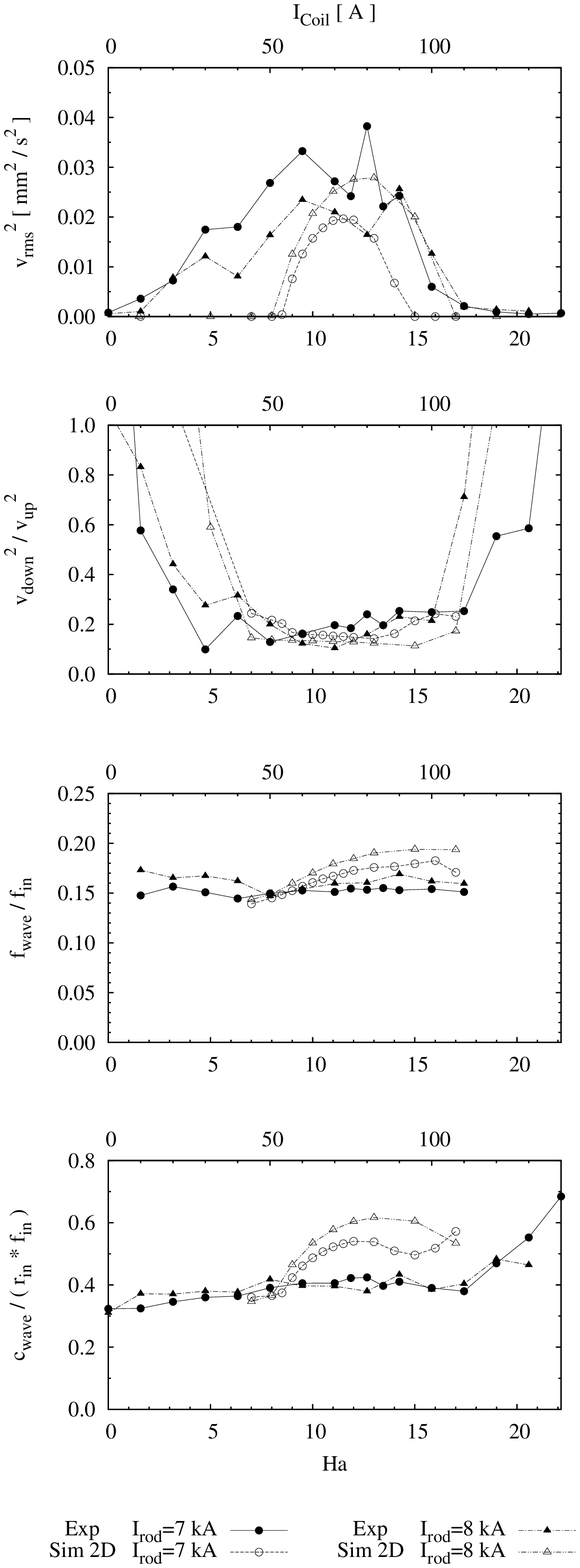}
\vspace{2mm}
\caption{Squared rms of $v_z(z,t)$, squared ratio of
maximum amplitude of downward to upward travelling waves, 
normalized 
frequency and normalized wave speed for 
$f_{\rm in}=0.06$ Hz, $\mu=0.27$,
$I_{\rm rod}=7000$ A and 8000 A,                               %!!!!
in depencence on $I_{\rm coil}$ (i.e. $Ha$).}
\label{Haanalyse}
\end{center}
\end{figure}


\begin{thebibliography}{99}
\bibitem{RUHOBUCH} G. R\"udiger and R. Hollerbach, {\it The Magnetic Universe}
(Wiley, Berlin 2004).
\bibitem{EARTH} D.J. Stevenson, Earth Planet. Sci. Lett. {\bf 208},
1 (2003).
\bibitem{SUN} M. Ossendrijver, Astron. Astrophys. Rev. {\bf 11}, 287 (2003).
\bibitem{GAL} A. Brandenburg and K. Subramanian, Phys. Rep. {\bf 417}, 1 (2005).
\bibitem{FLUCT} A.A. Schekochihin and 
S.C. Cowley, Phys. Plasmas {\bf 13}, 056501 (2006)
\bibitem{BAHA} S.A. Balbus and J. F. Hawley, Astrophys.\ J. {\bf 376},
 214 (1991).
\bibitem{VELIKHOV} E.P. Velikhov, Sov. Phys. JETP {\bf 36}, 995 (1959).
\bibitem{CHANDRA1} S. Chandrasekhar, Proc. Natl. Acad. Sci. {\bf 46}, 253 (1960).
\bibitem{TAYLER} R.J. Tayler, Mon. Not. R. Astron. Soc. {\bf 161}, 365 (1973).
\bibitem{SPRUIT1} H.C. Spruit, Astron. Astrophys. {\bf 381}, 923 (2002).
\bibitem{SPRUIT2} R. Moll, H.C. Spruit, and M. Obergaulinger, 
Astron. Astrophys. {\bf 492}, 621 (2008).
\bibitem{RMP} A. Gailitis, O. Lielausis, E. Platacis, G. Gerbeth, and
 F. Stefani, Rev.\ Mod.\ Phys.\ {\bf 74}, 973 (2002).
\bibitem{ZAMM} F. Stefani, A. Gailitis, and G. Gerbeth, ZAMM {\bf 88}, 930 (2008).
\bibitem{PRLRIGA} A. Gailitis et al., Phys. Rev. Lett. {\bf 84}, 4365 (2000)
\bibitem{KARLSRUHE} R. Stieglitz and U. M\"uller, Phys. Fluids {\bf 13}, 561 (2001).
\bibitem{MONCHAUX} R. Monchaux et al., Phys. Rev. Lett. {\bf 98}, 044502 (2007).
\bibitem{BERHANU} M. Berhanu et al., EPL {\bf 77}, 59001 (2007).
\bibitem{LATHROP} D.R. Sisan, N. Mujica, W.A. Tillotson, Y.M. Huang, 
 W. Dorland, A.B. Hassam, T.M. Antonsen, and D.P. Lathrop,
 Phys.\ Rev.\ Lett.\ {\bf 93}, 114502 (2004).
\bibitem{BURIN} H.T. Ji, M. Burin, E. Schartman, and J. Goodman,
Nature {\bf 444} 343 (2006).
\bibitem{ROACH} A. Roach, H. Ji, W. Liu, and J. Goodman, Bull. Amer. Phys. Soc.
{\bf 52}, No. 11,  BP8.00084 (2007).
\bibitem{HORU} R. Hollerbach and G. R\"udiger,
 Phys.\ Rev.\ Lett.\ {\bf 95}, 124501 (2005).
\bibitem{HORUAN} G. R\"udiger, R. Hollerbach, M. Schultz, and D.A.
 Shalybkov, Astron.\ Nachr.\ {\bf 326}, 409 (2005).
\bibitem{LIU1} W. Liu, J. Goodman, I. Herron, 
and H. Ji, Phys. Rev. E {\bf 74}, 056302 (2006).
\bibitem{PRIEDE1} J. Priede, I. Grants, and G. Gerbeth, Phys. Rev. E {\bf 75}, 047303 (2007).
\bibitem{JACEK1} J. Szklarski and G. R\"udiger, Astron. Nachr. {\bf 327}, 844 (2006).
\bibitem{LIU2} W. Liu, J. Goodman, and H. Ji, Phys. Rev E {\bf 76}, 016310 (2007).
\bibitem{JACEK2} J. Szklarski,  Astron. Nachr. {\bf 328}, 499 (2007).
\bibitem{JACEK3} J. Szklarski and G. R\"udiger, Phys. Rev. E {\bf 76}, 066308 (2007).
\bibitem{LAKHIN} V.P. Lakhin and E.P. Velikhov, Phys. Lett. A {\bf 369}, 98 (2007).
\bibitem{JACEK4} J. Szklarski and G. Gerbeth, Astron. Nachr. {\bf 329}, 667 (2008).
\bibitem{LIU3} W. Liu, Astrophys. J. {\bf 692}, 998 (2009).
\bibitem{PRIEDE2} J. Priede and G. Gerbeth, Phys. Rev E (in press); e-print  arXiv:0810.0386
\bibitem{KNOB1} E. Knobloch, Mon. Not. R. Astron. Soc. {\bf 255}, P25 (1992).
\bibitem{KNOB2} E. Knobloch, Phys. Fluids {\bf 8}, 1446 (1996).
\bibitem{HAWLEYBALBUS} J.F. Hawley and S.A. Balbus, Astrophys. J.
 {\bf 400}, 595 (1992).
\bibitem{RMPDISSIPATION}R. Krechetnikov and J.E. Marsden, Rev. Mod. Phys. {\bf 79}, 
519 (2007).
\bibitem{DEADZONES} N.J. Turner, T. Sano, and N. Dziourkevitch, Astrophys. J. {\bf 659},
729 (2007).
\bibitem{BALBUSPM} S.A. Balbus and P. Henri, Astrophys. J. {\bf 674}, 408 (2008).
\bibitem{HOLLERBACHRUEDUGERPRE} G. R\"udiger and R. Hollerbach, Phys. Rev. E. {\bf 76},
068301 (2007).
\bibitem{HUERRE} P. Huerre and P.A. Monkewitz, Ann. Rev. Fluid Mech. {\bf 22}, 473 (1990). 
\bibitem{CHOMAZ} J.M. Chomaz, Ann. Rev. Fluid Mech. {\bf 37}, 357 (2005).
\bibitem{GAILITISFREIBERGS} A. Gailitis and Y. Freibergs,
Magnetohydrodynamics {\bf 16}, 116 (1980).
\bibitem{PRL} F. Stefani, Th. Gundrum, G. Gerbeth, G. R\"udiger, 
M. Schultz, J. Szklarski, and R. Hollerbach, Phys. Rev. Lett. {\bf 97}, 184502 (2006).
\bibitem{APJL} G. R\"udiger, R. Hollerbach, F. Stefani, 
Th. Gundrum, G. Gerbeth, and R. Rosner, Astrophys. J. {\bf 649}, L145 (2006).
\bibitem{NJP} F. Stefani, Th. Gundrum, G. Gerbeth, G. R\"udiger, J. 
Szklarski, and R. Hollerbach, New J. Phys. {\bf 9}, 295 (2007).
\bibitem{AN} F. Stefani, G. Gerbeth, Th. Gundrum, J. Szklarski, 
G. R\"udiger, and R. Hollerbach, Astron. Nachr.  {\bf 329}, 652 (2008).
\bibitem{KAGEYAMA} A. Kageyama, H. Ji, J. Goodman, F. Chen, and
 E. Shoshan, J. Phys.\ Soc.\ Jpn.\ {\bf 73}, 2424 (2004).
\bibitem{PRIEDE3} J. Priede, e-print arXiv:0902.4896 
\bibitem{MAHYD} F. Stefani, G. Gerbeth, Th. Gundrum, J. Szklarski, 
G. R\"udiger, and R. Hollerbach,  e-print arXiv:0812.3790.
\bibitem{CHIFFAUDEL} N. Garnier and A. Chiffaudel, Phys. Rev. Lett. {\bf 86}, 75 (2001).
\bibitem{KNOBLOCHJULIEN} B. Jamroz, K. Julien, and E. Knobloch,
Physica Scripta {\bf T132}, 014027 (2008). 
\bibitem{CHANDRA2} S. Chandrasekhar, {\it Hydrodynamic and hydromagnetic instability}
(Dover, New York 1981).
\bibitem{RINCON} F. Rincon, G.I. Ogilvie, 
and M.R.E. Proctor, Phys. Rev. Lett. {\bf 98},
254502 (2007).
\bibitem{FOREST} E.J. Spence, K. Reuter, and C.B. Forest, e-print arXiv:0901.3406.
\end{thebibliography}
\end{document}